# Puzzle of the γ→α and other phase transitions in cerium

**A.V. Nikolaev[a], A.V. Tsvyashchenko[b]**

*We discuss recent research on the phase transitions in cerium under pressure, including new experiments on the γ→α transition, which indicate that it is a hidden structural phase transition.*

PACS: 61.50.Ks, 64.70.K-, 71.10.-w, 71.27.+a, 71.70.Jp, 76.80.+y, 81.30.-t;

[a]e-mail: alex_benik@yahoo.com

[b]e-mail: tsvyash@hppi.troitsk.ru

**Contents**

# 1. Introduction

Cerium was discovered in 1803 simultaneously in Sweden and Germany. It was named after the dwarf planet Ceres found two years earlier [1]. (Currently Ceres is the largest and most massive body in the asteroid belt, and contains almost a third (32%) of the belt's total mass.) Cerium is the most abundant (about 0.0046-0.0068 % of the Earth's crust by weight) and less expensive of the rare earth elements. Nowadays cerium is wide used for commercial applications. For example, in the form of cerium oxide it is added to fuel to reduce emissions and to glass and enamels to change their color, in the form of flint (iron-cerium alloy) it is used in lighters. Cerium oxide ($CeO_2$) is also an important component of glass polishing powders and phosphors used in screens and fluorescent lamps.

**Table 1.** Crystal structures of cerium allotropes.

| Phase | Crystal structure | Space symmetry | Lattice parameters (Å) | Conditions, Ref. |
|---|---|---|---|---|
| α | cubic | $Pn\bar{3}m$, $Pa\bar{3}$ (see Sec. 6 and 7) | $a = 4.824$ | $P = 0.81$ GPa, room $T$ [15] |
| β | double hexagonal close pack (dhcp) | $P6_3/mmc$ | $a = 3.6810$ $c = 11.857$ | $P = 0$, room $T$ [16] |
| γ | face centered cubic (fcc) | $Fm\bar{3}m$ | $a = 5.1610$ | $P = 0$, room $T$ [16] |
| δ | body centered cubic (bcc) | $Fm\bar{3}m$ | $a = 4.11$ | $P = 0$, $T = 1041$ K [2,4] |
| α' | α-U structure: C-centered (bases) orthorhombic | $Cmcm$ | $a = 3.0143(2)$ $b = 5.8935(3)$ $c = 5.1603(3)$ $y = 0.1014(2)$ | $P = 7.5$ ГПа, room $T$ [8] |
| α'' | monoclinic | $C2/m$ | $a = 5.813(2)$ $b = 3.145(1)$ $c = 5.612(2)$ $\beta = 113.10(2)°$ $(x,y,z)=(0.2800(5)),$ $0,0.2516(6))$ $c'=6.300$ $\beta'=55.02°$ | $P = 8.3$ ГПа, room $T$ [8] |
| ε | body centered tetragonal (bct) | $I4/mmm$ | $a = 2.92$ $b = 4.84$ | $P = 17.5$ ГПа, room $T$ [17] |

However, from the fundamental physics point of view cerium is the most interesting because of its unusual electronic structure which manifests itself in phase polymorphism of its condensed state [2]. Nowadays, seven allotropic phases of cerium are distinguished: α (cubic), β (double hexagonal close packed – dhcp [3]), γ (face centered cubic – fcc, $Fm3m$), δ (body centered cubic – bcc [4]), α' (C-type orthorhombic [5,6] or the α-U structure), α'' ($C2/m$ monoclinic [7,8]), ε (body centered tetragonal – bct, $I2/m$ [9,10]). (Table 1, 2 and Fig. 1). It is also worth noting that the density of the liquid phase is larger than the density of the bulk δ–phase. In addition, the liquid phase of cerium is stable in an unusually wide range of temperatures (2648 K) yielding only thorium. It is believed that plutonium, which also exhibits several allotropies - α, β, γ, δ, δ', (ζ), is the actinide analog of cerium [11].

**Table 2.** Nearest bond lengths in cerium phases.

| Phase | Bond lengths (Ce-Ce, Å) | Volume per Ce (Å$^3$) | Lattice parameters (Å) | Conditions |
|---|---|---|---|---|
| α | 3.411 | 28.065 | $a = 4.824$ | $P = 0.81$ GPa, room $T$ |
| β | 3.681 (6) 3.647 (6) | 34.784 | $a = 3.6810$ $c = 11.857$ | $P = 0$, room $T$ |
| γ | 3.649 | 34.367 | $a = 5.1610$ | $P = 0$, room $T$ |
| δ | 3.559 | 34.713 | $a = 4.11$ | $P = 0$, $T = 1041$ K |
| α' | 2.8435 (2), 3.0394 (4), 3.0143 (2), 3.3098 (4) | 22.918 | $a = 3.0143(2)$ $b = 5.8935(3)$ $c = 5.1603(3)$ | $P = 7.5$ GPa, room $T$ |
| α" | 2.9633 (1), 3.1142 (2), 3.145 (2), 3.3046 (4), 3.3525 (2), 3.3629 (1) | 23.593 | $a = 5.813(2)$ $b = 3.145(1)$ $c = 5.612(2)$ $\beta = 113.10(2)°$ | $P = 8.3$ GPa, room $T$ |
| ε | 2.92 (4) 3.181 (4) | 20.634 | $a = 2.92$ $b = 4.84$ | $P = 12$ GPa, room $T$ |

**Figure 1.**
Pressure-temperature phase diagram of cerium. Interphase α-β-γ boundaries are from Ref. [2], α-α'-α" boundaries (corresponding to the directions α→α', α→α" and α"→α') from Ref. [25], the boundary between α and ε phases from Ref. [26], the refined critical point of the γ→α transition is from Ref. [70]. The region of phase ambiguity corresponds to the mixture of phases (α, α', α" and ε), with the percentage depending on the $P$-$T$ trajectory leading to a $P$-$T$ point [25].

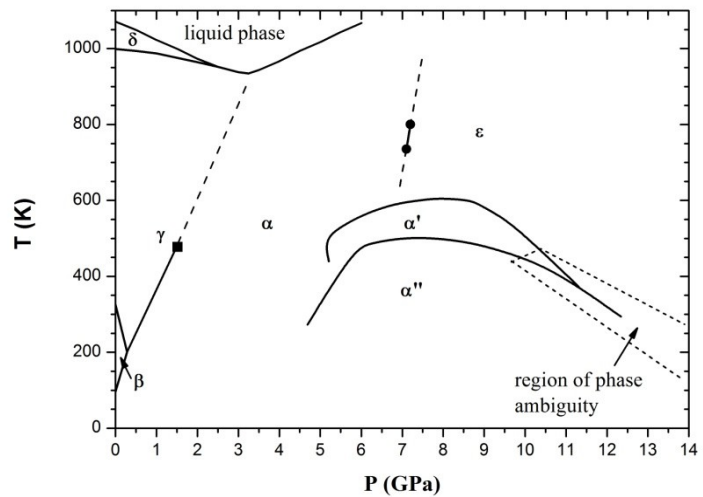

Among the phase transformations of cerium the most known transition is from γ-Ce to α-Ce [2]. Usually it is observed at room temperature under pressure of $P \approx 0.8$ GPa, but it also can be induced by temperature lowering at atmospheric pressure. In the latter case the transition goes through the intermediate β-phase [3], which is practically impossible to get rid of afterwards. For the first time the γ→α transformation was observed by Bridgman in 1927 [12]. However, the most famous property of the transition – its isostructureness – was established in x-ray experiments only in 1949 and 1950, first by applying pressure [13], and then by lowering temperature [14]. In both cases the same face centered structure for γ-Ce and α-Ce was reported. Thus both phases differ only by the numerical value of the fcc lattice constant (see Table 1). Since earlier two energy minima of the same structure had not been observed, the issue was considered of great scientific interest. Later however other isostructural phase transitions were found – for example, in $NpO_2$ ($T_c$ = 25.5 K) [19,18] and $YbInCu_4$ ($T_c$ = 42 K) [20] ($NpO_2$ is discussed below in section 6.6.1). The γ→α transition is accompanied by such a big volume change at room temperature (15%) that the term "volume collapse" was applied to this case (Fig. 2). This volume change sometimes is called unique, but it is not really so. For example, in going from graphite to diamond the volume decrease per carbon atom is much larger (57-69%). In that case we deal with *different covalent bonding* of carbon, which is well studied both theoretically and experimentally. Inspired by the carbon case we may suggest that in γ-Ce and α-Ce we are dealing with a *different manifestations of the metallic bond* of cerium. In particular, the metallic bond in γ-Ce can differ from the bond in α-Ce by its space orientation. In that case the γ→α transition will not be really isostructural. That will be a hidden structural transition closely connected with the symmetry change of the cerium electron density. This idea lies at the basis of the theory of quadrupolar ordering which is discussed in section 6.

**Figure 2.**

Volume change (per cerium atom) at the γ→α phase transition under pressure at room temperature. Experimental data are from Ref. [27].

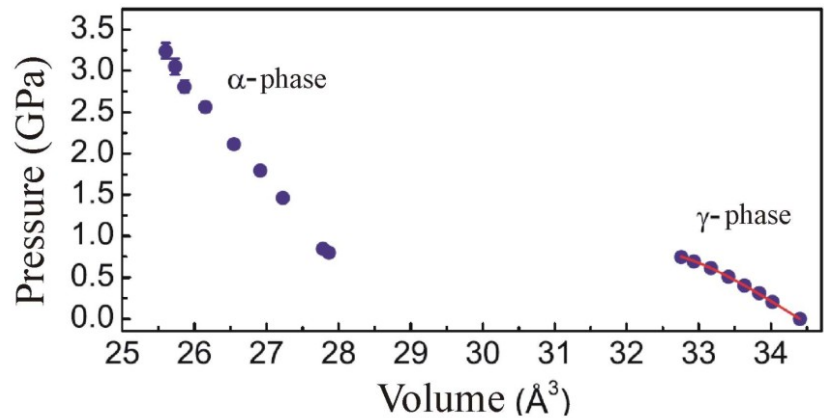

In addition to the γ→α phase transition a great scientific discussion was prompted by the problem of the identification of the phase in which cerium transforms from α-Ce under pressure [8]. As possible candidates the monoclinic structure (*I*2/m [10] or *C*2/m [7,8]) or base centered orthorhombic (the α-U structure) [5,6] were considered. Nowadays it is established that the symmetry of this phase depends on the method of production and sample history [8]. If a sample is prepared in the cold (low-temperature) synthesis it is transformed to the α"-phase of cerium, which is the monoclinic *C*2/m structure. If the sample is prepared at high temperature (and pressure) and cooled to room temperature, then the transition is from the α-cerium to the α'-phase with the α-uranium structure. In review the question is discussed in section 9.

The phase diagram of cerium (Fig. 1) is also constantly being discussed and refined, especially in the region of phases α, α', α" and ε, and transitions between them [21-26]. Phase boundaries between them were determined by measuring the electric resistivity of a sample at several *P-T* points. In the first study of Endo and Fujioka [21], the phase boundary between the α' (or α") and ε phases was found to form a straight line which extrapolated to the minimum of the cerium melting curve. These data were used in the reference book [22]. However, already in 1981 Antonova et al [23] found a dome-shaped curve of phase boundaries which was later confirmed by the work of Zhao and Holzapfel [24], and Tsiok and Khvostantsev [25]. In [25,26] the cerium phase diagram was studied in the region of high temperatures and pressures. There, the direct phase transition from the tetragonal phase (ε) to the cubic α-phase [25] and the transition in the opposite direction (т.е. α→ε) were found at pressures 7.1 GPa (T≈735 K) and 7.2 GPa (T≈800 K) [24,26] (see Fig.1). On the basis of these studies a suggestion on the existence of α-α'-ε triple point at 6.9 GPa and 600 K was made [26].

In this paper we consider the above mentioned problems and in particular, we discuss recent experimental data on the γ→α phase transition [27-29]. We also review other phase transitions under pressure paying much attention to the theoretical and computational methods.

## 2. Peculiarities of electronic structure of atomic cerium and the chemical bond of cerium dimer (Ce$_2$)

It is widely accepted that the phase variety of cerium with four valence electrons (4f 5d 6s$^2$) is caused by its single 4f electron. Indeed, cerium is virtually the first element in the D.I. Mendeleev Periodic Table in which an appreciable occupation of the 4f electron shell takes place. However, the direct participation of the 4f states in metal bonding is doubtful. In cerium properties of valence electrons, that is, the 4f electron from one side and 5d 6s$^2$ from the other are so different that they are grouped in two separate electron subsystems [36]. Three valence 5d 6s$^2$ electrons are usually considered as conventionally metallic, meaning that they form the typical chemical (metallic) bond whereas the core-like 4f electron remains mainly localized. In particular, this conclusion follows from the radial dependence of the partial electron density (Fig. 3). From the general theory of chemical bonding it follows that the optimal radius of the cerium atoms is found at a maximal value of its valence electron density. In accordance with this criterion we see (Fig. 3 and Table 3) that it is rather 5d and 6s electrons which are responsible for the metallic bond, because their electron density is maximal at $r_\gamma$ which is the contact radius (the radius of touching spheres) of γ-Ce. Meanwhile the 4f electron does not participate or almost does not participate in the process because its average radius $r_f$ is considerably smaller the characteristic bonding length ($r_f << r_\gamma$) (see Table 3). Below we will see that this scenario is fully supported by many electron relativistic calculation of the cerium dimer (Ce$_2$) [32,33].

**Figure 3.**
Partial electron density (4f, 5d, 6s) and valence electron density (5d 6s$^2$) in cerium (the relativistic DFT-LDA calculation of the cerium atom [185]). Radii $r_\gamma$, $r_\alpha$, and $r_{Ce2}$, are the radii of the contact spheres in γ-phase, α-phase and in cerium dimer (Ce$_2$), correspondingly.

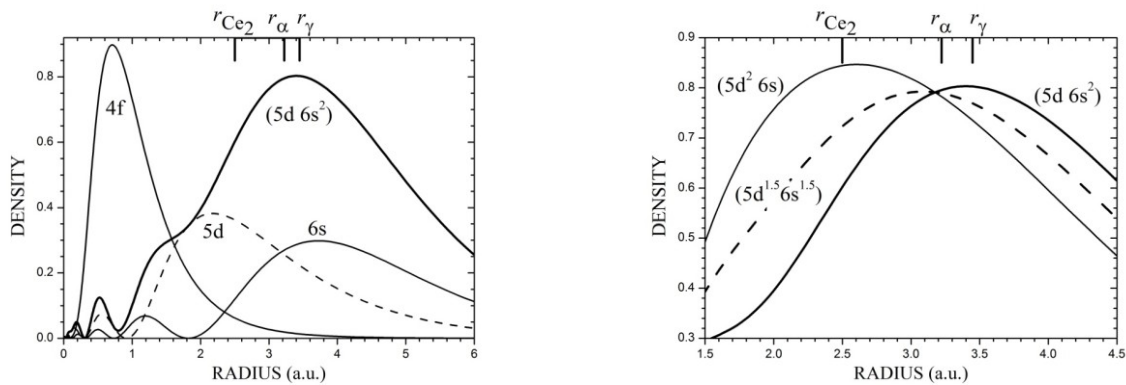

**Table 3.** Radial maxima ($\rho(r)r^2$) of atomic shells and valence electron density of cerium (Å). (The data are obtained from the relativistic atomic calculation of cerium within the density functional theory [107] in the local density approximation [185].) For comparison, the radii $r_c$ of the close contact spheres in solid state are: $r_\gamma$=1.825, $r_\alpha$=1.706, in cerium dimer $r_{Ce2}$=1.31-1.33 (Å).

|  | 4f5/2 | 5d3/2 | 6s1/2 | ($5d^1 6s^2$) | ($5d^{1.5} 6s^{1.5}$) | ($5d^2 6s^1$) |
|---|---|---|---|---|---|---|
| $r_{max}$ | 0.375 | 1.154 | 1.971 | 1.800 | 1.634 | 1.379 |

However, indirectly the 4f electron does affect the metallic bonding and it certainly cannot be considered as an ordinary core electron which simply screens the nuclear charge. Unlike other core electrons the 4f shell is open. This fundamental property implies 4f electron degrees of freedom which are absent in the case of a fully occupied electron shell. The completely filled electron shell leads to the spherically symmetric electron density and to the trivial spherically symmetric potential which screens the nuclear field. In atomic cerium because of the orbital degrees of freedom 4f states are grouped (coupled) with states of other valence electrons - 5d and 6s to minimize the resultant intrasite 4f-5d and 4f-6s Coulomb repulsions (Fig. 4 and Table 4). Thus already in the atomic cerium there exist rather strong 4f-5d and 4f-6s interactions which induce 4f-5d [37] and 4f-6s [38] correlations.

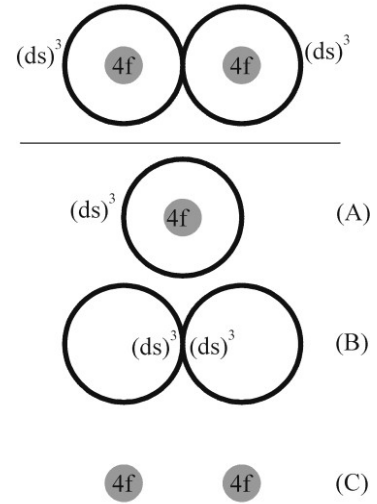

**Figure 4.**

Main electronic interactions between two neighboring cerium atoms in a crystal: (A) intra-atomic Coulomb repulsion ("Hund rules' coupling"), (B) chemical (metallic) bond between overlapping d- and s-electrons, (C) inter-site Coulomb 4f-4f interaction. Since the last interaction is the weakest it is responsible for the low-lying excitations.

These correlations can be found in the atomic spectrum of cerium (see Table 4). Its four valence electrons – 4f 5d $6s^2$ – are grouped to the $^1G_4$ many electron ground state in which the spin quantum number S = 0 whereas according to the first Hund rule S should be maximal [40]. In the atomic physics this is a rare case of first Hund rule *violation* [40]. The origin of the anomaly is discussed in detail in Ref. [39], where it is shown that in the $^1G_4$ state the quartic Fermi hole of forth order is formed which optimizes the energy of the 4f-5d Coulomb repulsion.

For the description of cerium in condensed state the authors of Refs. [41-43] model many electron effects of the 4f 5d $6s^2$ valence shell by introducing a spin polarization and an orbital term which artificially restore the first Hund rule for atoms of Ce. Taking into account that the first Hund's rule is not valid for the $^1G_4$ ground state of the cerium atom, the adequacy of the approach and its results is under question.

**Table 4.** Low-lying energy spectrum of the cerium atom.

| Leading configuration | term | J | g (Landé) | $\Delta E$ (eV), experiment [45] | $\Delta E$ (eV) [33] |
|---|---|---|---|---|---|
| 4f 5d 6s$^2$ | $^1$G | 4 | 0.945 | 0 | 0 |
| 4f 5d 6s$^2$ | $^3$F | 2 | 0.765 | 0.028 | 0.076 |
| | | 3 | 1.077 | 0.206 | 0.197 |
| | | 4 | 1.077 | 0.384 | 0.397 |
| 4f 5d 6s$^2$ | $^3$H | 4 | 0.890 | 0.159 | 0.188 |
| | | 5 | 1.032 | 0.274 | 0.305 |
| | | 6 | 1.160 | 0.493 | 0.539 |
| 4f 5d 6s$^2$ | $^3$G | 3 | 0.735 | 0.172 | 0.227 |
| | | 5 | 1.150 | 0.521 | 0.521 |
| 4f 5d$^2$ 6s | $^5$H | 3 | 0.600 | 0.294 | 0.299 |
| | | 4 | 0.986 | 0.302 | 0.340 |
| | | 6 | 1.166 | 0.589 | 0.631 |
| | | 7 | 1.237 | 0.719 | 0.752 |
| 4f 5d 6s$^2$ | $^1$D | 2 | 0.937 | 0.295 | 0.338 |
| 4f 5d$^2$ 6s | $^5$I | 4 | 0.666 | 0.396 | 0.602 |
| | | 5 | 0.907 | 0.467 | 0.690 |
| | | 6 | 1.117 | 0.552 | 0.797 |
| | | 7 | 1.216 | 0.659 | 0.921 |
| | | 8 | 1.250 | 0.844 | 1.066 |
| 4f$^2$ 6s$^2$ | $^3$H | 4 | 0.805 | 0.591 | |
| | | 5 | 1.035 | 0.774 | |
| | | 6 | 1.169 | 0.965 | |

Another peculiarity is the presence of the excited 4f 5d$^2$ 6s configuration (see $^5$H in Table 4), which dominates in the cerium atomic spectrum starting from 0.29 eV. Clearly, 4f 5d$^2$ 6s mixes up with 4f 5d 6s$^2$ in molecules and solids. Indeed potentially the 6s→5d transition can lead to a strengthening of chemical bonding because the 6s-shell becomes unoccupied and allows for the formation of the 6s-6s bonding between neighboring cerium atoms in molecules and solids. As shown in Table 3 and Fig. 3 the strengthening of metallic bonding results in an effective decrease of the characteristic radii and bond lengths. Interestingly, a partial increase of the d-orbital occupation was found in the Compton scattering experiment [44], so the discussed transfer indeed occurs at the γ→α phase transition in cerium. It is worth noting that on the basis of effective unit cell energy it was concluded that various solid phases differ from each other not only by crystal structure but also by their electronic structure [46]. Different electronic structure of γ and α cerium was also found in optical conductivity measurements [47].

It is instructive to study the peculiarities of chemical bond formation in the cerium dimer [34], because it is the simplest electron system and the simplest molecular bond binding two cerium atoms. Here, many electron calculations are implied where relativistic effects are taken into account. Calculations of that level become possible only recently [32,33]. According to their results, a triple bond is realized in Ce$_2$: the first bond is formed between 6s states of neighboring atoms, two others are due to 5d states. (If two cerium nuclei lie on the z-axis, the double d-bond is formed between the d-functions of two cerium atoms with the same azimuthal dependence ($m=\pm1$), that is between $\exp(-\varphi i)$ and $\exp(-\varphi i)$, $\exp(+\varphi i)$ and $\exp(-\varphi i)$.) Correspondingly, there are six valence electrons on this bond: $(6s\sigma_g)^2(5d\pi_u)^4$ (the $^1\Sigma_g^+$ many electron state). More importantly, at each cerium atom there remains one 4f-electron which does not participate in chemical bonding. The conclusion on inertness of 4f-states is not trivial. For example, in the uranium dimer (U$_2$) 5f-states behave differently and do participate in the formation of the quintuple chemical bond [48].

Although two remaining 4f electrons of cerium do not form a chemical bond, they play an important role in the low-lying excitation spectrum of Ce$_2$. In ab initio calculations [32,33] six low-lying many electron states ($^1\Sigma_g^+$, $^1\Sigma_u^-$, $^3\Sigma_g^-$, $^3\Sigma_u^+$, $^3 6_u$, $^1 6_g$) were found which are practically indistinguishable (0.01–0.07 eV). In [49] it is shown that the existence of such excitations is caused by 4f-4f interactions where the dominant one is the Coulomb repulsion of two localized 4f-electrons. Thus, in cerium the following scenario takes place (Fig 4). The chemical bond of Ce$_2$ is made of 5d and 6s electron states which form occupied molecular orbitals. The fixed bonding 5d and 6s orbitals through the Coulomb repulsion directly influence the space of the 4f states. An effective active space of the 4f states is narrowed to just two orbital f-states (with the azimuthal dependences $\exp(-3\varphi i)$ and $\exp(+3\varphi i)$). However, unlike other valence states (5d and 6s) the electron degrees of freedom of the localized f-electron are still not completely frozen. These electron degrees of freedom of nearly degenerate 4f-states are responsible for the low-lying molecular energy spectrum. It is conceivable that the same scenario takes place in the crystalline cerium.

### 3. γ→α phase transition in cerium (the review of models)

As we already discussed in the Introduction, the unusual feature of the γ→α phase transition is its isostructural nature discovered in 1949-50 [13,14]. In 1970-80 the focus was on different magnetic properties of the γ and α phases [2,50-52], and theoretical models were expected to explain this difference.

The magnetic susceptibility ($\chi_\gamma$) of γ-cerium follows the Curie-Weiss law (Θ = -50 or -9 K, $p_{eff}$ = 2.4-2.52 μ$_B$) [50,51] with an effective magnetic moment close to the value $p_{eff}$ = 2.54 μ$_B$ for a single free f-electron. The magnetic susceptibility of α-cerium shows only weak dependence on temperature, so that α-Ce is characterized as a Pauli paramagnet [52,53]. However, the minimal value of $\chi_\alpha$ (at $T \approx 50$ K) is 4.5 larger than the value deduced from the electronic specific heat measurements [53]. $\chi_\alpha$ demonstrates a slow increase with temperature when $T > 60$ K and steep increase on decreasing temperature below $T < 20$ K (usually it is accounted for by β-impurities). The data are conventionally explained by the presence of nearly free f-electrons in the γ-phase which becomes bound in the α-phase. However, this is a simplistic interpretation. Indeed, in Sec. 2 we have seen that even in the cerium atom the f-electron is not free because through the Coulomb repulsion it interacts with d- and s- valence electrons. The strength of this interaction is of the order of 1 eV (see Table 4), so the fact that the effective magnetic moment coincides with that of a free f-electron is probably coincidental [49]. As in the case of dimer, in condensed cerium there can be low-lying many electron 4f excitations. These excitations may be magnetic (with a non-zero magnetic moment) and nonmagnetic (with zero magnetic moment). Their characteristic splittings are of the order of 10-100 K, and currently it is practically impossible to predict their energy positions with confidence. (Nevertheless, in some cases it can be done, see Ref. [49].)

The γ→α transition is possibly a record-holder for the number of models suggested for its description. First, it was considered that the localized f-electron which is present in the γ-phase becomes a part of the metallic (6s5d) band in the α-phase. This is the promotional model [54-56], which at that time was linked with the cerium valence instability [57,58]. Thus the transition occurs according to the following scenario:

$$(\gamma\text{-Ce})\ 4f^1 5d^1 6s^2 \rightarrow 4f^0 5d^2 6s^2\ (\alpha\text{-Ce}),$$

from which it follows that the γ and α phases differ by the number of f-electrons. This conclusion was checked in the positron annihilation experiments [59,60] and in Compton scattering experiments [44]. In both cases nearly equal occupation of the f-states ($n_f \approx 1$) was established and the promotional mechanism was concluded irrelevant. This fact is supported by an accurate theoretical consideration. Indeed, the 4f→5d (4f→6s) transition is very energy consuming. It can be approximated by 4 eV from the experimentally measured transitions (4f $5d^2$ ($4H_{7/2}$) → $5d^3$ in $Ce^{1+}$ [45]). In addition, the calculations of the electron band structure of γ and α phases by the density functional method (DFT) in the local density approximation (LDA) have shown that the number of occupied f-states in both phases is almost equal to one per cerium atom [61]. Recently, the conclusion concerning the number and nature of f-states was even strengthen: the inelastic neutron scattering experiments indicate that the 4f-electron remains localized at the γ→α phase transition [62,63].

A detailed analysis on the partial occupations of valence states in γ and α phases given in the Compton scattering measurements [44] is particularly worth mentioning. It has turned out that the number of 4f-electrons in the α-phase close to one, and even a small change of the occupation number leads to a worsening of the correspondence between the experiment and the model. At the same time the occupation numbers of d- and s-states do change: in the α-phase the number of d-states is increased by 0.6 which perfectly fits the scenario of the chemical bonding which we discussed in Sec. 2.

In 1974 having analyzed the thermodynamic properties of cerium, Johansson put forward a Mott-like (localization-delocalization) model for the 4f-electron subsystem at the γ→α phase transformation [36]. Now the number of 4f electrons remains constant in both phases but the change concerns their nature: in α-phase the electrons are band-like, while in γ-phase they are localized. Two fcc phases (γ and α) then correspond to two minima of the free energy which develop within the same space symmetry ($Fm\bar{3}m$). Later, the theory was confirmed by model band structure calculations of cerium [41-43,64].

The description of the γ→α transition in cerium and other metals (for example, in actinides [65,66]) within the dilemma of localization-delocalization of f-states [36,41-43,64] have become wide spread. However, the explanation of the physical and chemical properties within the concept remains essentially one-electronic. In many electron approach this contradistinction is not really relevant. Indeed, the localized atomic or the delocalized Bloch functions are only characteristics of different basic sets. From the Bloch states by a corresponding linear transformation one can obtain the localized atomic functions and vice versa, whereas a correct solution of the electronic problem should not depend on the choice of the basis set. One can say of course that the Bloch functions are preferable because they explicitly take into account the translation symmetry of a crystal. However, the Bloch functions are symmetry adapted only for the one electron problem. In the realistic many electron case the translation symmetry is *exact* only if all crystal electrons are subjected to a translation. The situation here is completely analogous to the many electron atomic one. In the atomic problem the many electron Hamiltonian is invariant in respect to the rotation operations which act on all electrons of the atom. In case of only one electron rotation it is wrong. The same holds for the crystal translations in the many electron case.

There is another approach called the Kondo volume collapse model which competes with the theory of Mott-like transition in cerium [30,31,67]. From its name it is clear that the model exploits a large volume change at the transition and the Kondo mechanism [68] for demagnetization of magnetic moments. In the Kondo model the 4f electron in a first approximation stays localized in the γ- and α-phases, but the interaction between the 4f-electrons and band electrons at the Fermi level is much more intense in the α-phase. Quantitatively that implies that in the γ-phase the Kondo temperature is small (50-100 K) while in the α-phase it is relatively large (1000-2000 K). In addition, the so called hybridization between 4f and band states gives rise to the formation of the singlet ground state which is separated from the magnetic states by an energy gap of $k_B T_K$. Thus, the last feature to be explained is the appearance of two minima of the same fcc lattice. It is worth noting that such type of behavior has not been observed in the molecular structures. Molecules often have close energy minima, but they always correspond to different structures (conformations). Therefore, in order to describe two minima the Kondo model always exploits a nonlinear empirical dependence. In the first formulation that was the dependence of the Kondo temperature on the cerium unit cell volume [30]. In recent formulation it uses a dependence of an effective Debye temperature [70].

Another shortcoming of the model (as well as the model of localization-delocalization of the 4f-electron density) is that it is effective only for one unique phase transition (γ→α). The other phase changes in cerium then should be explained by the traditional means. Therefore, one phase transition (γ→α) is singled out from the whole set of phase transformations, which is not completely logical.

Two abovementioned models (Mott-like and Kondo-like for 4f-states) have been actively developed during the last 20 years and from the technical point of view they become more and more complicated (see Sec. 5.1). Nowadays, the Kondo effect is described within the dynamical mean-field theory (DMFT) (see Sec. 5.2) and has lost the characteristics of empirical approach. However, even today there is no consensus on the unambiguous approval of just one of these models [70,71]. We finally note that recent inelastic neutron scattering measurements of α-Ce (in a sample doped by 10 at. % Sc) [62] have found that its magnetic form-factor differs appreciably from the calculated one [42] where the 4f-states are considered itinerant (delocalized) (Fig. 5). Thus, according to Ref. [62], the 4f-electrons remain localized even in α-Ce. This conclusion is in full agreement with works [32,33] on the many electron calculations of the cerium dimer from which it also follows that the 4f-electrons do not participate in chemical bonding (see Sec. 2).

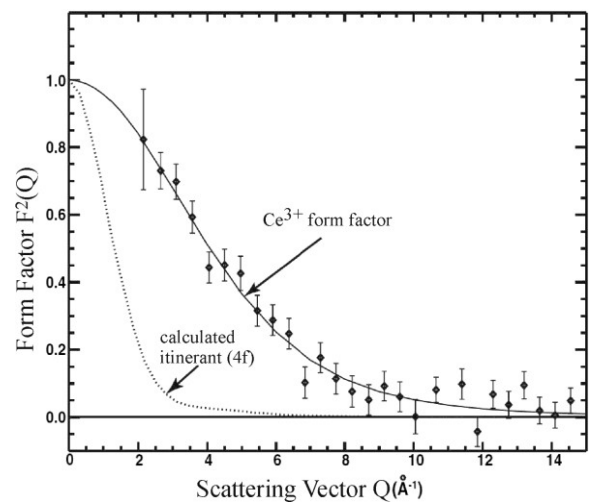

**Figure 5.**

The magnetic form-factor $F^2(Q)$ in α-Ce (in $Ce_{0.9}Sc_{0.1}$) obtained by high energy neutron inelastic scattering measurements [62]. The solid curve through the experimental data corresponds to the $Ce^{3+}$ form factor [72]. The dotted line is a calculated theoretical curve for the itinerant 4f electrons [42] (LMTO-LDA method).

It is worth mentioning that both in the Kondo model and the localization-delocalization of 4f electron subsystem model the isostructural nature of the phase transition has never been questioned and has always been considered as a well established fact. However, there is another approach [73-75] which treats the phase change as a *hidden structural* transformation (see Sec. 6). In 2010 the approach for the first time received an experimental support [28,29].

## 4. Elastic anomalies and the role of lattice in the γ→α transition

The elastic phenomena at the γ→α transition were first observed and reported in 1961 by Voronov et al. [78] and in 2009 accurate ultrasonic measurements were carried out by Decremps et al. [69]. Over last decade the role of lattice through the transition has been much discussed starting from the work of Jeong et al. [27] where it was emphasized that the vibrational contribution to the entropy change per atom is about half of the total entropy change ($1.5k_B$ [2]). In [184] however it was found that the conclusion does not hold for the γ→α transition in $Ce_{0.9}Th_{0.1}$ alloy, which in many aspects is similar to the transition in elemental cerium. In 2011 Krisch et al. [80] reported only a moderate lattice contribution to the entropy change ($\Delta S_{vib}^{\gamma-\alpha} = 0.33k_B$) in pristine cerium. This value is about half of that of Ref. [27]. In this Section we consider the active role of lattice in the γ→α transition which manifests itself through a softening of some phonon branches of γ-cerium [76,77] and softening of the bulk modulus [78,79,69]. The elastic anomalies can be quantitatively explained within the quadrupolar model [186].

### 4.1 Softening of the bulk modulus at the γ→α transition and other elastic phenomena

The elastic phenomena at the γ→α transition were first observed and reported by Voronov et al. [78]. At room temperature in polycrystalline cerium they monitored the bulk modulus ($B$), the shear modulus ($G = c_{44}$), an effective Debye temperature ($\Theta_D$) and Poisson's ratio as functions of applied pressure in ultrasound measurements. As pressure increased above 0.4 GPa and approached the onset of the γ→α transition they found that $B$ and Poisson's ratio fast decreased while $G$ slowly increased and the Debye temperature $\Theta_D$ remained nearly constant. Upon completing the transition to the α-phase all elastic moduli and the Debye temperature increased step-wise, after which they continue to rise slowly. This behavior was farther confirmed in more precise measurements [69,79]. In [79] experimental data were collected up to the pressure of 9 GPa and thus covered not only γ→α but also the following α→α' (or α→α" [8]) phase transition. Highly accurate ultrasonic measurements on ultra pure polycrystalline cerium up to 1 GPa were collected recently by Decremps et al., Ref. [69].

The phonon frequencies of γ-cerium single crystals along the [100], [110], [111], [0ξ1] high-symmetry directions of the Brillouin zone were studied by Stassis et al. [76,77] (Fig. 6). Comparison of the phonon dispersion curves with those of thorium [81] shows that the spectrum of γ-cerium is in general softer (i.e. its vibration frequencies are smaller) than one expects from the Lindemann criterion for melting. The softening is more pronounced for the longitudinal (L) modes and the T[111] and $T_1$[110] transverse modes, and less pronounced for the $T_2$[110] and T[100] branches (see Fig. 6). The T[100] mode almost corresponds to that of thorium (no softening) [81]. Therefore, the softening is absent for those branches (T[100] and $T_2$[110]) which in the long wavelength limit ($\vec{q} \to \upsilon$) have no connection with the elastic constants $c_{11}$ and $c_{12}$. The dominant effect – softening of the phonon spectrum of γ-Ce – is directly related to the decrease of its bulk modulus $B = (c_{11} + 2c_{12})/3$.

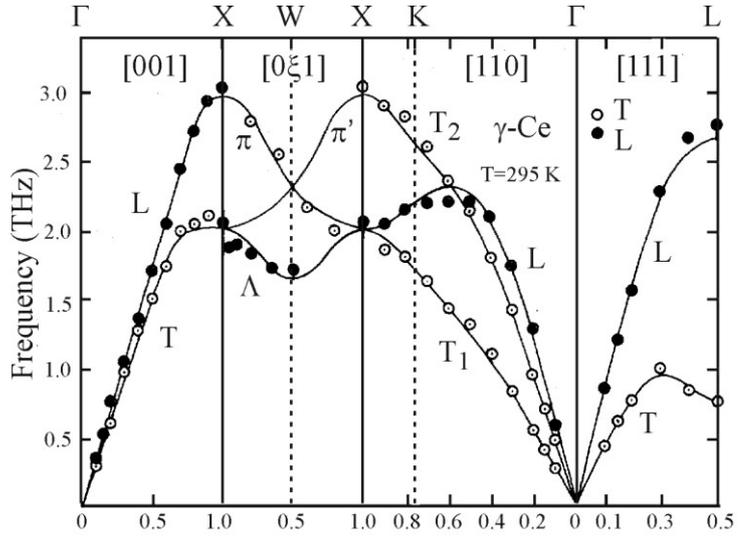

**Figure 6.**

Phonon dispersion curves of γ-Ce (295 K) extracted from the inelastic neutron scattering measurements [76]. The solid lines were obtained by fitting the data in the force constant model (the Born-von Kármán model with 8 nearest coordination shells).

Note that very low frequencies along the Γ-L line (the T[111] curve) are caused by another (martensitic) phase transition, γ→β (fcc→dhcp) which occurs at 260 K (see Sec. 8). This conclusion is supported by the anomalous behavior of the phonon modes around the L: $(2\pi/a)(1/2,1/2,1/2)$ point of the Brillouin zone [77]. The frequency at the L point drops as temperature decreases from 0.94 THz (875 K) до 0.82 THz (295 K), while for the other modes of γ-cerium the reverse effect is observed [77]. The same soft mode (≈0.82 THz) at the L point was also found in lanthanum (T = 660 K), where the transition of the same type occurs: fcc→dhcp. (The fcc→dhcp transformation involves four transverse waves propagating along the [111] direction with reduced three vectors $(2\pi/a)(\xi,\xi,\xi)$, where $\xi = 1/4, 1/3$ and $1/2$ [82].)

From the disperse curves of γ-cerium Stassis et al. found the atomic force constants of the Born-von Kárman model (Table 5), which fully determine the phonon spectrum, that is: the effective Debye temperature $\Theta_D(T)$ ($\Theta_D$=119–135 K), and the elastic constants: $c_{11}$=2.41, $c_{12}$=1.02 and $c_{44}$=1.94, in $10^{10}$ N m$^{-2}$ or $10^{11}$ dynes cm$^{-2}$. Also, they concluded that there was no well-defined low-energy crystal field excitation in γ-Ce. The obtained values of the elastic constants imply large crystal anisotropy. In particular, the shear elastic constant $c_{44}$ is three times larger than the quantity

$$c' = (c_{11} - c_{12})/2 , \qquad (1)$$

that is $A=c_{44}/c'$=2.8. It has turned out that in the high temperature δ-phase (bcc) the anisotropy is even higher, $A$=5.6 [82]. The phonon spectrum of δ-Ce [82] and its peculiarities are discussed in detail in Sec. 8.

The dependence of phonon dispersion on pressure at room temperature was recently studied by high energy resolution inelastic x-ray scattering [80]. In particular, in this work for the first time the phonon dispersion curves of α-Ce were obtained at $P$ = 0.8 and 2.5 GPa. Krisch et al. have found that dispersion curves change mainly along the [110] and [111] directions with the most noticeable difference in the vicinity of the X point. The Grüneisen parameters $\gamma_q$ for modes T[001], T$_2$[110] and L[110] become negative around the X point, which on one side implies a phase instability and on the other points out its special role.

Jeong et al. [27] have studied the behavior of the bulk modulus in polycrystalline cerium samples in the proximity of the γ→α transition obtaining it by finite differences of the $P$-$V$ data, $B = -VdP/dV \approx -V\Delta P/\Delta V$ (Fig. 7). Pronounced softening of the bulk modulus of γ-phase as the pressure approaches the onset of the phase transition is an indication of increasing elastic instability of lattice, and the discontinuity of $B_T$ at the transition implies that the transition is of the first order.

**Table 5.**

Force constants $\Phi_{\alpha\beta}(\vec{R}_n)$ (in n/m or $10^3$ dynes/cm) of γ-Ce in the Born-von Kármán model with 8 nearest coordination shells obtained by fitting the phonon spectrum [76]. N is the number of atoms in shell, $\alpha\,(\beta) = x, y, z$.

| $\vec{R}_n$ | N | xx | yy | zz | xy | xz | yz |
|---|---|---|---|---|---|---|---|
| a(1/2,1/2,0) | 12 | 4.373 | | -0.226 | 4.580 | | |
| a(1,0,0) | 6 | -2.356 | 0.077 | | | | |
| a(1,1/2,1/2) | 24 | 0.206 | 0.317 | | | -0.050 | -0.055 |
| a(1,1,0) | 12 | 0.123 | | 0.011 | 0.151 | | |
| a(3/2,1/2,0) | 24 | -0.053 | -0.099 | -0.104 | 0.019 | | |
| a(1,1,1) | 8 | -0.332 | | | | | -0.219 |
| a(3/2,1,1/2) | 48 | 0.106 | -0.114 | 0.026 | -0.007 | 0.076 | 0.005 |
| a(2,0,0) | 6 | 0.001 | 0.222 | | | | |

The bulk modulus in the γ-phase can be fit by the polynomial dependence $B_T \propto |P - P_c|^\alpha$ ($P_c$ = 0.83 GPa, $\alpha$ = 0.46). The obtained isothermal bulk modulus $B_T$ is in good agreement with the adiabatic bulk modulus $B_S$ obtained from the ultrasound measurements by Voronov et al. [78,79]. In the cubic phase the bulk modulus is given as

$$B = \frac{1}{3}(3c_{11} - 4c' + P),$$

where $c'$ is the shear modulus, and $P$ is the applied hydrostatic pressure. Since in a first approximation $c'$ is independent of pressure [78,79,69], the softening of the bulk modulus is directly related with the softening of $c_{11}$. The conclusion is confirmed by the softening of the measured longitudinal sound velocity with increasing pressure [78,79,69] and the transverse phonon frequencies at normal pressure [76,69]. The anomalous softening of $c_{11}$ can be quantitatively explained within the quadrupolar model [186].

**4.2 Mean square atomic displacements and the lattice contribution to the entropy change**

Isotropic mean square thermal displacements of atoms in γ-Ce and α-Ce were studied in Refs. [27,83,104] where Jeong et al. used a new experimental technique. Traditionally phonon spectra are obtained from the inelastic neutron scattering or from a high energy resolution inelastic x-ray scattering often obtained in synchrotron radiation measurements. In both cases a single crystal is required whereas powder samples are considered inappropriate for the purpose. However, the study of the γ→α transition in cerium single crystal involves some serious technical problems (see Sec. 7). Therefore, the alternative approach based on the atomic pair distribution function analysis obtained by means of high energy synchrotron and pulsed neutron sources on powder samples is of much importance [84]. The pair distribution function is a Fourier transform of the diffraction spectrum into real space. From the pair distribution function (the peak positions and their widths) one can obtain the mean square thermal displacements of cerium nuclei $\langle u_{iso}^2 \rangle$, which then are used for further analysis. In Ref. [27] the mean square thermal displacement $\langle u_{iso}^2 \rangle$ of a cerium atom in the crystal lattice in γ- and α-phases was studied as a function of temperature at 0.412 and 0.527 GPa and a function of pressure at 300 K.

The lattice contribution to the entropy change is estimated from
$$\Delta S_{vib}^{\gamma-\alpha} \equiv \Delta S_{vib}^{\gamma} - \Delta S_{vib}^{\alpha} \approx 3k_B \ln(\Theta_D^{\alpha} / \Theta_D^{\gamma}).$$

In [27] very different Debye temperatures were found for the two phases ($\Theta_D^{\gamma}$=104(3) K and $\Theta_D^{\alpha}$=133(3) K) and
$$\Delta S_{vib}^{\gamma-\alpha} = (0.75 \pm 0.15) k_B.$$

Thus, according to [27] the estimated vibrational entropy change per atom in the γ→α transition, $\Delta S_{vib}^{\gamma-\alpha}$, is about 50% of the total entropy change, $\Delta S_{tot}^{\gamma-\alpha}$ = 1.5 $k_B$, obtained from the latent heat and the Clausius-Clapeyron relation [2]:
$$dP/dT = \Delta S_{tot}^{\gamma-\alpha} / \Delta V^{\gamma-\alpha}.$$

The phonon dispersion of cerium as a function of pressure was recently studied by high energy resolution inelastic x-ray scattering [80]. The authors have found that the contribution to the γ→α transition entropy ($\Delta S_{tot}^{\gamma-\alpha}$ =0.33 $k_B$), is only 22% of the total entropy change. For completeness we quote the Debye temperature and $\langle u_{iso}^2 \rangle$ obtained in [80] just before and after the phase transition: $\Theta_D^{\gamma}$=122 K, $\Theta_D^{\alpha}$=138 K; $\langle u_{iso}^2 \rangle$=0.0206 Å² in γ-Ce and 0.0168 Å² in α-Ce. On the basis of highly accurate ultrasonic measurements on ultrapure polycrystalline cerium Decremps et al. [69] estimated the lattice contribution to the entropy change at 15%.

On the other hand, neutron diffraction study on the γ−α phase transition in $Ce_{0.9}Th_{0.1}$ alloy [104] implies that the vibrational entropy change there is not significant.

**4.3 Interpretation of experimental data**

In [27] it is pointed out that the dependence of the mean square thermal displacement $\langle u^2 \rangle$ on temperature or pressure and the behavior of the bulk modulus $B_T$ in the γ→α transition can be described within the well known model [85,86], where the energy contribution (order parameter-strain field coupling) has the following functional form:
$$\Delta F_{vib} = \lambda \varepsilon \rho^2. \quad (2)$$
Here ε is a characteristic lattice strain (a component of the strain tensor), and $\rho$ is an order parameter amplitude. The effect of lattice softening preceding a structural phase transition is a universal feature of the coupling (2). The physical mechanism responsible for the elastic instability and a renormalization of the bulk modulus were first studied in the theoretical works of Wagner and Swift [85] and Bergman and

Halperin [86]. These models and the coupling (2) itself indicate that the lattice displacement (volume collapse) is a secondary order parameter.

Note that the contribution of the same type was obtained within the quadrupolar model (see Eqs (6), (7) below and Sec. 6.2, 6.3). For more details, see Ref. [186].

## 5. Models for the electronic structure of γ-Ce and α-Ce without space symmetry change

Cerium is a unique element in the field of electron structure of correlated metals, with its γ→α transition being a testing ground for the implementation of new theoretical approaches in the computational solid state physics. This results in a great number of studies in which its electronic structure is modeled [41-43,64,65,71,90-103], starting with the first DFT-LDA calculation [90] and ending with recent works [96-102] within the dynamic mean field theory (DMFT) [104,105]. Some of the studies along with principal approximations and computed characteristics are given in Table 6.

It is very important to note that in all these calculations [41-43,64,65,71,90-103] the isostructural nature of the transition is taken for granted. Meanwhile, new recently emerged experimental data indicate that the γ→α phase change is a hidden structural transition (see Sec. 6 and 7) [28,29,73-75,37].

### 5.1 Electron band structure calculations of cerium

The pioneer calculation of cerium is done by Glötzel [90] within the local spin density approximation of DFT. In the spin-polarized treatment corresponding to the unrestricted Hartree-Fock method [106], the electron density is split in spin-up and spin-down components, while for the exchange interaction an empirical expression [107] of the free electron gas is used. In this approach the description of γ-Ce is unsatisfactory, since the ground state is ferromagnetic with a wrong equilibrium volume. The subsequent studies followed the path of more and more sophisticated computation schemes which incorporated uncontrolled empirical approaches. For example, in studies [41-43] a so called orbital polarization was introduced. That is an energy term added to an effective one electron Hamiltonian, which simulates three Hund's rules for cerium. This procedure ignores the fact that in the cerium atom the first Hund rule is violated (see Sec. 2). In works [92-95] band structure calculations were corrected to explicitly exclude the self interaction of the localized 4f states of cerium. (The electronic self-interaction [108] is an artifact of the local density approximation of DFT which appears as a result of the mismatch between the exact Coulomb and an approximate exchange interaction.)

Correlation effects caused by the many electron Coulomb repulsion are conventionally reduced to the Hubbard interaction and a multiplet splitting of the localized 4f levels [109] (the so called "Hubbard I" or "LDA++" approximation). It is notable however, that all correlation effects in that approach are limited to the 4f states. The band $(spd)^3$ electrons of cerium are treated in a one electron approximation (as an electronic background or so called an "electronic thermostat"). This is a serious approximation by itself. In Ref. [38] it is shown that by averaging the single site s-f interactions we retain only even terms for cubic crystals and thus exclude important odd electron density fluctuations which appear in the exchange between s- and f-electrons (octupole interactions in terms of multipole expansion).

**Table 6.** Electronic structure of γ-Ce and α-Ce in various band structure models.

| Ref. | authors | year | main approximations | calculated values |
|---|---|---|---|---|
| [90] | Glötzel D. | 1978 | spin-polarized muffin-tin LMTO (DFT-LDA) | equilibrium lattice constants, bulk modulus, magnetic susceptibility of γ-Ce, La, Th |
| [61] | Pickett W E, Freeman A J, Koelling D D | 1981 | non-muffin-tin LAPW (DFT-LDA) | magnetic susceptibility, occupation of 4f-states |
| [41] | Eriksson O, Brooks MSS, Johansson B | 1990 | spin and orbital polarization in LMTO (DFT-LDA), simulation of Hund rules | equilibrium volumes in γ-Ce and α-Ce |
| [92] | Svane A | 1994 | self-interaction correction for 4f-states, spin-polarization in DFT-LDA | equilibrium volumes in γ-Ce and α-Ce |
| [93] | Szotek Z, Temmerman WM, and Winter H | 1994 | self-interaction correction for 4f-states, spin-polarization in DFT-LDA | equilibrium volumes and magnetic moments in γ-Ce and α-Ce |
| [64] | Johansson B et al. | 1995 | Mott-like transition in DFT-LDA | γ→α phase diagram, critical point, free energy contributions |
| [91] | Söderlind P et al. | 1995 | generalized gradient approximation (GGA), spin-polarization in DFT-LDA, full potential | equilibrium volume, bulk modulus, equation of state for α-Ce |
| [95] | Laegsgaard J and Svane A | 1999 | self-interaction correction for 4f-states, magnetic impurity in DFT-LDA | equilibrium volumes and magnetic moments in γ-Ce and α-Ce, temperature effects |
| [96] | Zölfl M B et al. | 2001 | dynamic mean-field theory (DMFT) with DFT-LDA | equilibrium volumes, photoemission spectra of γ-Ce and α-Ce |
| [97] | Held K, McMahan A K, and Scalettar R T | 2001 | dynamic mean-field theory (DMFT) with DFT-LDA, quantum Monte-Carlo | equilibrium volumes and magnetic moments in γ-Ce and α-Ce |
| [99] | Haule K et al. | 2005 | dynamic mean-field theory (DMFT) with DFT-LDA | optical conductivity for γ-Ce and α-Ce and comparison with experiment [47] |
| [89] | Amadon B et al. | 2006 | dynamic mean-field theory (DMFT) with DFT-LDA for LMTO | entropy change at the γ→α transition, equilibrium volumes, photoemission spectra |

**5.2 The dynamic mean field theory (DMFT)**

Last decade for the description of the peculiarities of the γ→α phase transition, methods of the dynamic mean field theory [104,105] (DMFT) have been applied [96-102]. DMFT is an approximation to the electronic task in a crystal lattice when the electron problem with many degrees of freedom is reduced to a single impurity task with essentially fewer degrees of freedom. The approximation becomes exact in the hypothetical limit $z \rightarrow \infty$, where $z$ is the number of nearest neighbors. (In practice $z \propto$ 2-12.) The nearest neighbors thus are substituted with a one-electron "thermostat". It is believed that DMFT describes the electronic situation better than the Mott-like and Kondo models.

The DMFT accounts for the hybridization between the localized f electron states and also between the delocalized spd- and localized f-electrons. First, the problem is solved in an effective one electron approximation which gives computed electron bands formed by the spdf-states. Then the Coulomb interaction in the form of the Hubbard repulsion $U\approx6$ eV comes into play and a certain realization of DMFT is applied. As a DMFT solver one can use the non-crossing approximation (NCA), Ref. [96], or quantum Monte-Carlo (QMC) and Hubbard-I, Ref. [98]. The main finding then is that at zero temperature the energy minimum corresponds to α-Ce, but when temperature is elevated to 0.14 eV, the minimum is shifted to a larger volume which is interpreted as the γ-phase of Ce.

Therefore, DMFT offers the following scenario for the γ→α phase transition. In a large unit cell volume (γ-Ce) the f-electron spectrum is split by the Hubbard repulsion which results in the appearance of local magnetic moments at cerium sites. With decreasing volume a quasi-particle at the Fermi energy is formed (the Abrikosov-Saule resonance) which leads to the disappearance of the magnetic moment and an entropy decrease. The temperature evolution of the quasi-particle peak can be explained by the difference in the Kondo temperature, which in α-phase ($T_{K,\alpha}$) is much larger than in the γ-phase ($T_{K,\gamma}$). In Ref. [96] $T_{K,\alpha}=$ 1000 K, $T_{K,\gamma}\approx$30 K, in Ref. [98] $T_{K,\alpha}=$ 2100 K, $T_{K,\gamma}<$ 650 K, the estimations of the Kondo temperature from the experimental data are $T_{K,\alpha}=$ 945 K, $T_{K,\gamma}=$95 K (electron spectroscopy [110]) and $T_{K,\alpha}=$ 1800-2000 K, $T_{K,\gamma}=$60 K (neutron spectroscopy [111]). Although the Kondo temperature in γ-Ce is very small DMFT still considers it as a strongly correlated solid. Some differences in parameters and results of Refs. [96,98] are explained by different methods of solving the single impurity problem. Interestingly, the authors of Ref. [97] found some similarity between DMFT and the Mott-like approach.

DMFT as any other method is not free from serious approximations. One of them is the one electron thermostat serving as a background for 4f-electron correlations. It implies that the 4f states are artificially singled out from the other electron states. Also, the employed one-electron functions are obtained within DFT-LDA, which is known to be an uncontrolled approximation. The other group of approximations is related to the parameter $1/z$ which is supposed to be small. This parameter formally prevents DMFT from being applied to molecules but it would be very instructive to study how this approximation works for well known test molecules (for example, in the case of cerium dimer [34]). In comparison with other "classical" many electron methods [32,33] well known in quantum chemistry, DMFT is a novel approach and therefore its errors and drawbacks are virtually unknown. Such particularities could be revealed very clearly if DMFT is compared with classic quantum chemical approach.

### 5.3 Attempts to explain the elastic anomalies and lattice contribution in the γ→α transition

As we already discussed in Sec. 4, both ultrasonic measurements [78,79,69] and x-ray [80] and neutron diffraction [27] have shown that the cerium lattice in the γ→α phase transition plays an active role: the bulk modulus of γ-Ce is softening as the pressure approaches the phase transition point.

In recent studies [70] and [89] attempts have been undertaken to comprehend the role of lattice [27] in the framework of the empirical Kondo model and the dynamic mean field theory, correspondingly. In Ref. [89] it has been even claimed that the α→γ phase transformation is driven by entropy change.

Note that in contrast to the theories of structural phase transitions, elastic anomalies are not intrinsic to the Kondo scenario (see Sec. 6). In addition, in recent study [80] it has been found that Grüneisen parameters $\gamma_q$ for a number of phonon modes become negative at the X point of the Brillouin zone. It is worth noting that the X point of the Brillouin zone is a very special one in the quadrupolar model, the orientational density mode at the X point drives the γ→α phase change, see Sec. 6.2-6.5.

## 6. γ→α transformation as a hidden structural phase transition

In Refs. [73-75,37] a distinctive approach is presented, which considers the γ→α phase transition as a hidden structural transformation. Note that its intrinsic feature is softening of bulk modulus and elastic anomalies [69,78,80] (Sec. 4). Recently, for the first time it was supported by experiments [28,29], using the method of time-differential perturbed angular γγ-correlations of the $^{111}$Cd probe nuclei (Sec. 7).

### 6.1 Thermodynamic analysis of Eliashberg and Capellmann

First doubts on the isostructural nature of the γ→α phase transitions were expressed in the work of Eliashberg and Capellmann [73]. Their main idea is that in cerium instead of the critical point between the γ- and α-phases there exists a "critical point of second order transitions" (the definition of Landau and Lifshitz, see Chap. 150 [112]), which later was called the tricritical point. In Ref. [112] it is further noted that such a tricritical point "in a sense is analogous to the usual critical point".

Its properties follow from the expansion of the Gibbs free energy (or the chemical potential, $\mu$) in terms of the order parameter amplitude $u$:

$$\mu(P,T,u) = \mu_0(P,T) + A(P,T)u^2 + B(P,T)u^4 + C(P,T)u^6.$$

At the tricritical point we have $A_c = 0$, $B_c = 0$, $C_c = 0$. The main property of the point is that the line of transitions of the first order defined by

$$B < 0, \; 4AC = B^2,$$

continues beyond the tricritical point as a line of transitions of the second order given by

$$B > 0, \; A = 0.$$

The conclusion that the γ-α phase boundary continues beyond the tricritical point as a line of transitions of the second order was completely new. In particular, it explains the minimum of the cerium melting curve which approximately lies on the line.

Analyzing further experimental thermodynamic data, the authors of Ref. [73] were able to reconstruct phase transition boundaries, the melting curve, the change of compressibility and entropy, which were in good agreement with experimental values.

Probably the most important conclusion which follows the treatment is that the space symmetry of the α-phase should be lower than the symmetry of the γ-phase. In [73] a distorted face centered cubic lattice with at least two nonequivalent atoms in the primitive unit cell was suggested. Eliashberg and Capellmann ascribed the phase instability to the soft transverse mode at the $L$ point of the Brillouin zone [76,77].

The study of Eliashberg and Capellmann was brought to the attention of experimentalists and prompted updated and more refined x-ray measurements for the α-phase [27,70,113]. However, the subsequent x-ray diffraction studies [27,70,113] have not revealed a distorted fcc structure in α-Ce, although in [70] a pronounced softening of the bulk modulus on the suggested line of transitions of the second order has been emphasized (Fig. 8).

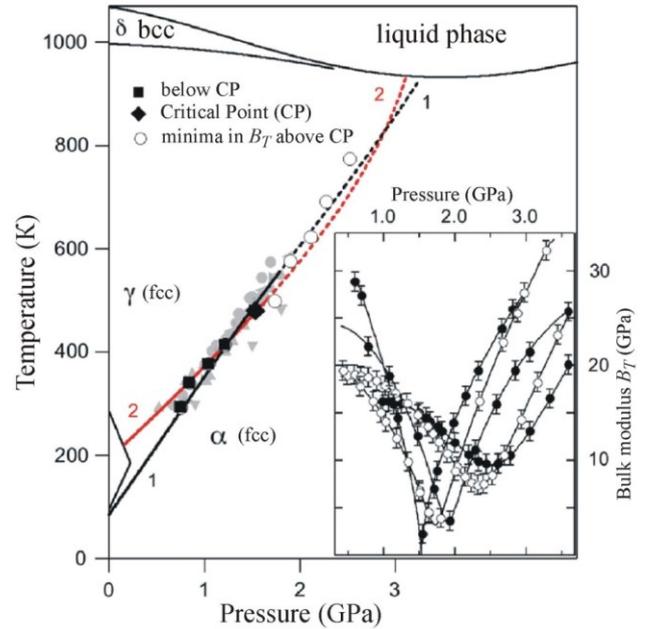

**Figure 8.**
The line of the bulk modulus minima in the cerium phase diagram continues beyond the critical point [70]. The black squares are taken from Ref. [70], the grey symbols are measurements of other authors. The open circles are bulk modulus minima above the critical point. Insert: the dependence of the bulk modulus at the following temperatures: 481, 500, 577, 692, и 775 K.

The idea on the nonisomorphic nature of the γ→α transition was independently formulated in Refs. [74,75,37], where a quadrupolar ordering of the electron density in the α-phase was considered as the driving force for the phase transformation. Although the quadrupolar model can be viewed as a continuation of the approach of Eliashberg and Capellmann, already in the first work [74] it was claimed that there is no distortions of the fcc lattice of cerium nuclei, and all structural changes are associated with electronic density of the valence electrons.

## 6.2 Quadrupolar model of the γ→α transition

In Refs. [74,75,37] the γ→α phase transition is considered to be structural, albeit with a very special space symmetry lowering. The main peculiarity is that the suggested symmetry change concerns the electron density but conserves the fcc lattice of cerium nuclei in the α-phase. We recall that usually it is the lattice distortion which indicates crystal symmetry lowering. Its absence according to the authors explains why the structural nature of the γ→α phase transformation has not been found and detected earlier. In that case we speak of a very specific or rather a hidden space symmetry lowering.

The quadrupolar model can be considered in broad and narrow sense. In the most general sense it implies a specific symmetry change which masks the structural phase transition. In the narrow sense the quadrupolar model is a physical model which takes into account only some important interactions. In that case the model is not complete because it neglects other interactions. In Ref. [74] the intersite Coulomb interaction between f-electrons was considered, in Ref. [75] the intersite Coulomb repulsion of s-, d-, f-electron densities were introduced and in Ref. [37] two electron (f- and d-) intrasite and intersite repulsions were accounted for. In the last case the model includes correlation effects which are omitted in the standard band structure calculations.

The simplest description is in the model of interacting f-electrons. As discussed in Sec. 2, the low-lying energy spectrum of the cerium dimer is caused by 4f-electron interactions with the Coulomb repulsion as the most important contribution. The Coulomb repulsion between f-electrons belonging to different sites is *the weakest* among the others (see Fig. 4), but that is why the interaction is responsible for the lowest excitations of the many electron spectrum (see [49]). Thus, it can be assumed that the driving force of phase transitions is the Coulomb repulsion of the 4f electrons localized at cerium sites because a change of their electron density costs little energy. This scenario has been studied in Ref. [74]. The Coulomb interaction between the 4f-electrons can be written in terms of the double multipole expansion,

$$U^{ff} = \frac{1}{2} \sum_{n,n'}{}' \sum_{L,L'} \rho_L^F(n) \upsilon_{L,L'}(n-n') \rho_{L'}^F(n'), \qquad (3)$$

Here $L$ stands for $(l,\tau)$, where $l$ is the angular index of the multipole expansion and $\tau=(\Gamma,k)$. $\Gamma$ refers to an irreducible representation of the cubic point group, while $k$ labels rows of $\Gamma$ [114]. The quantity $\rho_L^F(n)$ is the operator of the multipole electron density at the cerium site $n$, while $\upsilon_{L,L'}(n-n')$ is the matrix of multipole interaction. Then the authors retain two important groups of terms in Eq. (3). These are the quadrupole interaction ($U_{QQ}^{ff}$, $l=l'=2$) and a crystal field contribution ($l=4$, $l'=0$). Having performed the Fourier transfer, $U^{ff}$ can be rewritten through Fourier transforms $\rho_Q^F(\vec{q})$ and $\upsilon_{Q\tau,Q\tau'}(\vec{q})$ ($\tau$=1-5). The 5x5 $\upsilon_{Q\tau,Q\tau'}(\vec{q})$ quadrupole matrix has negative eigenvalues in some points of the Brillouin zone, which implies an effective attraction between 4f electrons. In particular, a large negative value ($-\lambda^X$) has been found at the $X$ point of the Brillouin zone [$\vec{q}_X = 2\pi/a(1,0,0)$]. It has two degenerate eigenvector which can be expressed through two components of the $T_{2g}$ symmetry [74]. This effective attraction leads to a phase instability at 86 K which results in the appearance of the quadrupole electron density and concomitant symmetry lowering. Since the space group of $q_X$ involves three rays: $2\pi/a(1,0,0)$, $2\pi/a(0,1,0)$, $2\pi/a(0,0,1)$, its electronic mode has six components. Condensation of three out of six density components at $q_X$ enables the $Fm\bar{3}m \to Pa\bar{3}$ structural phase transition [114-116]. Thus γ-Ce is identified as a disordered phase (with the order parameter amplitude $\rho$=0) with the $Fm\bar{3}m$ space symmetry, whereas α-Ce as the quadrupolar ordered phase of cubic symmetry but with different space group – $Pa\bar{3}$.

Condensation of a single component of one ray (for example at $\vec{q}_X$) implies that in real space this component changes sign in going from one crystallographic plane (perpendicular to the $x$ axis) to another. Condensation of all three components means that the sign is changed in going from one plane to another along the $x$, $y$ and $z$ axes. Such structures are called triple-$q$-antiferroquadrupolar or 3-$q$-AFQ. A close examination shows that the $Pa\bar{3}$ structure is described by four different sublattices of the simple cubic lattice (Fig. 9).

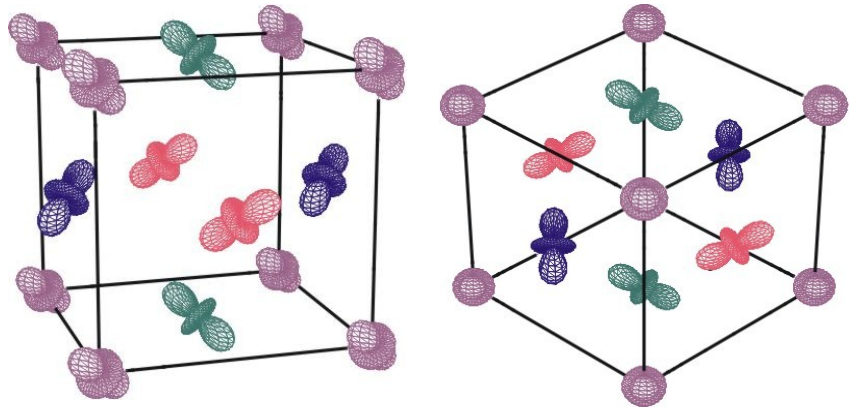

**Figure 9.**
The triple-$q$-antiferroquadrupolar structure (3-$q$-AFQ) proposed for α-Ce in Ref. [74]. The quadrupolar ($l$=2) functions represent the electron density component of the four (4f+5d6s$^2$) valence electrons. Right panel: the view along one of the main cube diagonal – [111] demonstrating the trigonal point site symmetry ($S_6$). (Figure from Ref. [28].)

The cerium atoms belonging to the same sublattice are completely equivalent, while cerium atoms of different sublattices are not. The four sublattices $n_p$ contain the following sites (which in the γ-phase were completely equivalent): $(0,0,0)$, $(a/2)(0,1,1)$, $(a/2)(1,0,1)$, $(a/2)(1,1,0)$. The symmetry lowering implies that in the ordered phase at each cerium site there is only one three-fold axis of symmetry ($C_3$), which simultaneously is a main cube diagonal. If the [111] direction is taken as the new $z'$ axis, the angular component of the quadrupolar density is described by the function $S_{[111]}(\theta',\varphi') = Y_\ell^{m=0}(\theta',\varphi')$, which is aligned along the $z'$-axis. In the $x, y, z$ coordinate system the angular function is given by

$$S_{[111]}(\theta,\varphi) = \frac{1}{\sqrt{3}}\left(Y_2^{1,s}(\theta,\varphi) + Y_2^{1,c}(\theta,\varphi) + Y_2^{2,s}(\theta,\varphi)\right).$$

We recall that in the fcc structure there are four three-fold axes of symmetry (the four main cube diagonals) $[\rho_1,\rho_2,\rho_3]$: [111], [-1-11], [1-1-1], [-11-1]. Thus, in the ordered (α-Ce) phase there remains only one out of these four axes for each of the four sublattices (see Fig 9). The angular quadrupolar function at any site can be expressed in the following form:

$$S(\theta,\varphi) = \rho\left(e^{i\vec{q}_y \cdot \vec{R}(n)} Y_2^{1,s}(\theta,\varphi) + e^{i\vec{q}_z \cdot \vec{R}(n)} Y_2^{1,c}(\theta,\varphi) + e^{i\vec{q}_x \cdot \vec{R}(n)} Y_2^{2,s}(\theta,\varphi)\right),$$

where $Y_2^{1,s}(\theta,\varphi)$, $Y_2^{1,c}(\theta,\varphi)$, $Y_2^{2,s}(\theta,\varphi)$ (proportional to $yz, zx, xy$) are real spherical harmonics [114], and $\vec{R}(n)$ is the radius vector of the site $n$. The quadrupole order involves all nonspherical states (i.e. p, d and f), so that in general we speak of the local quadrupole density component of all four valence cerium electrons [37,75]. This component is noticeably weaker than the spherically symmetric electronic density present both in γ- and α-phases and formed by all 58 cerium electrons.

### 6.3 Volume change and bulk modulus softening in the quadrupolar model

The quadrupolar model was considered for a deformable crystal lattice [74]. The lattice then adjusts to the change of electron density caused by the appearance of quadrupole components, which is a secondary effect. Without going into detail, we note that by taking into account derivatives of Coulomb interactions one obtains an additional contribution $U^{QQT}$ (quadrupole-quadrupole-translation) in the Hamiltonian of deformable crystal lattice, which is bilinear in terms of the order parameter amplitude $\rho$ (quadrupole) and linear in terms of lattice displacements $u_\nu(\vec{q})$ (translation) [74]. This contribution is analogous to the well known rotation-rotation-translation term, $U^{RRT}$, which has been thoroughly studied in the theory of phase transitions of molecular crystals [117,118]. In particular, the compression of the fullerite lattice (in the $C_{60}$ crystal) at the $Fm\bar{3}m \to Pa\bar{3}$ phase transition [119,120] was considered in Ref. [117]. In cerium within the quadrupolar model the situation then is quite analogous to the model of [117].

As discussed in Sec. 6.2, the phase transition itself is driven by the condensation of the order parameter at the $X$ point of the Brillouin zone, $\vec{q}_X$. For the deformable lattice the condensation leads to the following contribution to the free energy (per atom):

$$F^{QQT}[\rho,\varepsilon] = -2a\Lambda\rho^2\left(\varepsilon_{xx} + \varepsilon_{yy} + \varepsilon_{zz}\right) \quad (4)$$

where $\varepsilon_{\nu,\nu}$ are the longitudinal strains, $\rho$ is the quadrupolar order parameter amplitude (see Sec. 6.2), $a$ is the lattice constant and $\Lambda$ is a parameter related to derivatives of the Coulomb interaction between quadrupole density components. Note that a special role of phonon modes at the $X$ point of the Brillouin zone was emphasized in recent inelastic x-ray scattering experiments [80]. Combining $F^{QQT}[\rho,\varepsilon]$ with the elastic lattice contribution $F^{TT}[\varepsilon]$ quadratic in terms of $\varepsilon_{\nu\nu}$ and with the free energy [74], and minimizing the resultant expression in respect to the strain tensor components, we obtain

$$\varepsilon_{xx} = \varepsilon_{yy} = \varepsilon_{zz} = 8a^{-2}\Lambda\kappa_L\rho^2 < 0 \quad (7)$$

where $\kappa_L = \left(c_{11}^0 + 2c_{12}^0\right)^{-1}$ is the bare lattice compressibility, while $c_{11}^0$ and $c_{12}^0$ are elastic constants. Since $\Lambda < 0$, Eq. (7) results in a homogenous lattice contraction in the ordered α-phase and a concomitant discontinuous decrease of the cubic lattice constant. If only f-electrons are considered the change of the lattice constant is small ($\Delta a$=-0.002 Å). It is increased by 4.4 if s- and d- electrons are included in the model [75], but the estimation does not take into account the bulk modulus softening (see Sec. 6.5).

It is also well known [85,86,121] that the $U^{RRT}$ interaction leads to a softening of the elastic constant

$$c_{11} = \frac{c_{11}^0}{1 + v^2 \chi / c_{11}^0}.$$

Here $c_{11}^0$ is the bare elastic constant in the absence of the interaction (4), $v$ is the corresponding interaction constant, while $\chi$ is the static susceptibility which is proportional to the quadrupole specific heat. In fact, this is a quadrupole analog of the compressible Ising model [85,86,121]. Thus, the softening of the bulk modulus [27,70] and the longitudinal sound speed [78] at the γ→α transition reflects the specific heat increase (in analogy with the deformable Ising lattice [85,86]). A detailed description of this effect can be found in recent theoretical study, Ref. [186].

Finally, we note that in Ref. [117] the Clausius-Clapeyron equation has been derived and it has been shown that the transition temperature $T_1$ is a linear function of external pressure $P$ with a positive coefficient of the proportionality. The corresponding expression for cerium is given by

$$\frac{dT_1}{dP_1} = 4a^{-2}\Lambda\kappa_L x^{(2)},$$

where $x^{(2)}$ is a one-particle expectation value. This expression accounts for the linear dependence of the γ/α boundary in the $PT$ diagram of cerium [2].

Therefore, the quadrupolar model of the γ-α transition is able to explain qualitatively many nontrivial effects: the volume decrease, the softening of the bulk modulus and the linear pressure increase of the γ/α boundary in the $PT$ diagram although the quantitative agreement is not so good.

### 6.4 Does the nuclear subsystem of α-Ce retain the fcc lattice in the quadrupole model?

The conclusion that fcc symmetry of nuclear subsystem is conserved at the γ→α transition was formulated in Refs. [37,74,75]. An alternative opinion on this problem presented in referee reports of Yu. A. Uspenski to this review is based on the following arguments: "Consider the electron density of the α-phase in the form of $\rho(\mathbf{r})=\rho_0(\mathbf{r})+\delta\rho_Q(\mathbf{r})$, where $\rho_0(\mathbf{r})$ is a part of the electron density having the same symmetry as in the γ-phase, while $\delta\rho_Q$ is a perturbation of the density, lowering the symmetry of the electron subsystem. The Coulomb potential (Hartree's potential) associated with the electrons in the α-phase will be $V(\mathbf{r})=V_0(\mathbf{r})+ \delta V_Q (\mathbf{r})$, where the potentials $V_0$ and $\delta V_Q$ are due to $\rho_0$ and $\delta\rho_Q$, correspondingly. The potential perturbation $\delta V_Q(\mathbf{r})$ should have the same low symmetry as $\delta\rho_Q(\mathbf{r})$. The electric field associated with $\delta V_Q(\mathbf{r})$ will act on all charges in the system including electrons and nuclei. Under the action of this field the nuclei will be displaced from the equilibrium positions and their symmetry will be lowered. Therefore, the symmetry of the electronic and nuclear subsystems should be the same." "The nuclear displacements can be small and difficult for experimental determination but principally nonzero."

In order to examine the situation, we consider the full potential in the form of multipole expansion. (The detailed multipole expressions for nonspherical crystal potential are discussed in Refs. [122,123]). Consider the cerium nucleus at the origin of the coordinate system and a main cube diagonal as the z-axis. Then according to the quadrupolar model the full potential $V(\mathbf{r})$ in the α-Ce has the quadrupole ($l=2$) and monopole ($l=0$) contributions. The monopole component is present also in γ-Ce. From the equilibrium condition for γ-Ce (a nucleus is in the potential minimum) we obtain:

$$V_0(\vec{r}) = v_0(r,\theta,\varphi) = -C_0 r^2 \qquad (C_0 > 0) \qquad (6)$$

In α-Ce there is also the quadupole density contribution

$$\delta\rho_Q(r,\theta,\varphi) = \rho_Q(r) Y_\ell^{m=0}(\theta,\varphi),$$

where $\rho_Q(r)$ is the radial function. The potential corresponding to this density contribution can be found by means of multipole expansion [75,122,123]:

$$\delta V_Q(\vec{r}) = \delta v_Q(r,\theta,\varphi) = \frac{4\pi}{5}\left(\frac{q_2(r)}{r^3} + r^2 q'_2(r)\right) Y_\ell^{m=0}(\theta,\varphi). \qquad (7)$$

where

$$q_2(r) = \int_0^r \rho_Q(r') r'^4 \, dr' \qquad \text{and} \qquad q'_2(r) = \int_r^R \rho_Q(r') r'^{-1} \, dr'. \qquad (8)$$

Here $q_2(r)$ describes a quadrupole charge, which is situated inside a sphere of radius r, while $q'_2(r)$ is an effective quadrupole potential due to the electron density outside the sphere. Notice that there are no divergencies at the point $r = 0$. That is because the quadrupole component can be formed by p-, d- and f-electrons ($l=1, 2, 3$), the wave function of which in the vicinity of zero behaves as $r^l$, and the density as $r^{2l}$. Therefore, in the region of $r\approx 0$ the first term can be omitted (since it gives a small dependence of the $r^4$ type). The function $q'_2(r)$ at $r\approx 0$ can be estimated from above: $q'_2(r) \approx q'_2(r=0) = q'_2$. Then for the potential (7) we get

$$\delta V_Q(r,\theta,\varphi) = C_2 r^2 Y_\ell^{m=0}(\theta,\varphi), \qquad (9)$$

where $C_2 = \frac{4\pi}{5} q'_2$. Notice that from the multipole expansion it follows that the contribution from the distant crystal region (the whole crystal except the cerium unit cell at the origin of the coordinate system) will also be of the same functional form (9). Thus this contribution can be accounted for by changing correspondingly the parameter $C_2$.

In α-Ce the full potential is $V(\mathbf{r}) = V_0(\mathbf{r}) + \delta V(\mathbf{r})$, from which we infer that the minimum at the point $r=0$ is conserved if the condition

$$C_0 > |C_2| \max\left|Y_\ell^{m=0}(\theta,\varphi)\right| = |C_2| \frac{1}{2}\sqrt{\frac{5}{\pi}}$$

is satisfied. This condition holds because the quadrupole component formed by four valence electrons is appreciably smaller than the monopole component formed by all electrons (58 electrons), i.e. $C_0 \gg |C_2|$. Therefore, our opinion is that in the α-phase the fcc lattice of cerium nuclei is conserved.

In connection with this discussion it is worth noting experimental results on $NpO_2$ [132], where the situation is somewhat analogous to the γ→α transition in cerium (see Sec. 6.5 and 6.6.1 for more details). In the low temperature ordered phase there appears a quadrupole component of electron density on the neptunium atoms forming the fcc sublattice, which symmetry (3-q-AFQ) is very close to the one predicted by the quadupole model. The quadrupole ordering in $NpO_2$ however does not lead to an experimentally observable deformation of the fcc sublattice of neptunium [18,132].

**6.5 Discussion of the quadrupolar model**

The quadrupolar model does predict a volume change in the γ→α transition, but if only the 4f-electron repulsion is taken into account [74] the numerical estimation is an order of magnitude smaller than the experimental value. However, one should take into account the following. First of all, as discussed in Sec. 4.1 and 4.2 on approaching the γ→α phase boundary one observes softening of certain phonon modes and a substantial decrease (see Figs 7 and 8) of the bulk modulus [27,70,76-78]. Such strong softening inevitably leads to the correspondent increase of lattice compression. Also, as discussed in Sec. 2 (see Fig. 4) the phase transition affects the conduction electrons which are responsible for metallic bonding. Note that the interaction between the 4f-electrons should be considered rather as a precursor one [75,37], which prompts a chain of electron reconstructions of 5d- and 6s- states responsible for the metallic bond in cerium (see Sec. 2). One should also keep in mind that a large volume change occurs only in a limited area of the phase diagram of cerium, specifically in the region of low temperatures and pressures where the influence of metallic 5d- and 6s- states is probably more pronounced, while in the vicinity of the critical point and below it a negligible change of volume is observed. It is conceivable that in that area only f-states are effective.

Note also that according to Eliashberg and Capellmann [73] (see Sec. 5.1) there is no true critical point. Since the symmetries of two phases differ, the interphase boundary continues up to the melting curve (see Fig. 1). The experimental information on the phase transition in this area (i.e. above the tricritical point and below the melting curve minimum) is dubious. Indeed in Fig. 8 (Ref. [70]) the bulk modulus demonstrates a characteristic V-shaped dependence versus pressure which implies a pronounced softening at the curve minimum even at the temperature 775 K and pressure 2.5 GPa, although *this point* (Fig. 1) *is situated far from the tricritical point and close to the melting curve minimum*. Therefore, the dash line can be an interphase boundary of transitions of the second order (as assumed in Ref. [73]) or transitions of the first order with small order parameter discontinuities [74].

The quadrupolar model can describe the crystal field, the phase transition, the symmetry lowering and the lattice displacements on equal footing. However, in the mean field approximation [74] the magnetic susceptibility of a single 4f electron localized at a cerium site is bound to follow the Curie (or Curie-Weiss) law. In practice, as discussed in Sec. 2 the 4f-electron through the multipole Coulomb repulsion always interacts with 5d- and 6s- electrons at the same site (the so called Hund's interaction) [37,38], so that even in the γ-phase 4f-, 5d- and 6s- states are correlated (see Table 4). Taking this effect into account one can explain the change in the magnetic moment of the ground state [37,38,124]. To calculate this effect with confidence is next to impossible because it requires the computation of the many electron energy spectrum with a very high accuracy.

Of course, the disappearance of magnetic moments at the γ→α phase transition can be described by Kondo-type models [107]. However, symmetry lowering considered in the quadrupolar model in principle allows us to obtain this effect solely from the group-symmetry arguments [29,37,38,124]. This mechanism is possible if as a result of symmetry lowering the magnetic state of γ-Ce splits in few sublevels and the lowest nonmagnetic sublevel (singlet) then becomes the ground state of the α-phase [38,124].

Earlier we have already emphasized that although according to the quadrupolar model the γ→α phase transition is structural with the $Fm\bar{3}m \to Pa\bar{3}$ symmetry lowering, experimentally it looks differently. The main reason is that before and after the phase transformation the cerium nuclei form the fcc lattice which hinders the experimental identification of true symmetry change. Meanwhile, the $Pa\bar{3}$ space symmetry is often found in molecular crystals. For example, a phase of such symmetry is observed in molecular ortho-hydrogen [125], in the $C_{60}$ fullerite [126,127], in molecular nitrogen (α-phase) [128]. In fact, the quadrupolar model [74] was inspired by the theory [129,130] of the $Fm\bar{3}m \to Pa\bar{3}$ phase transition in the $C_{60}$ fullerite. However, in all molecular solids the transformation to the $Pa\bar{3}$-structure implies nuclei ordering and therefore it can be identified relatively easy. In metals with cubic lattices it is not so and the recognition of the 3-*q*-AFQ structure becomes a nontrivial experimental problem. Nevertheless hidden structural phase transitions directly related with a quadrupole density ordering have been found in some rare earth compounds (see Sec. 6.5.1). Among them the most remarkable example is perhaps $NpO_2$ [19,131,132]. For a long time $NpO_2$ had been thought to undergo an isostructural phase transition at 25.5 K. When however, the oxide was studied by the resonant x-ray spectroscopy it was found that the low-temperature phase has a hidden order related with the quadrupole ordering of electron density [132]. It turned out that the phase transition in $NpO_2$ is of the $Fm\bar{3}m \to Pn\bar{3}m$ type, and the symmetry of the ordered phase ($Pn\bar{3}m$) is very close [133] to $Pa\bar{3}$ suggested in the quadrupolar model [74] (Fig. 10 and Sec. 7). Thus, in $NpO_2$ one observed the effect which was expected to occur in pristine crystalline cerium. In principle additional reflections from the cerium electron density could be found similarly in x-ray diffraction experiments with synchrotron radiation source especially since a new resonant x-ray scattering technique – RXS – has been developed recently [134-136]. However, recent experimental data in support of the quadrupole order in α-Ce has been obtained differently by the nuclear method of perturbed angular correlations [28,29] (Sec. 7).

**Figure 10.**

Quadrupolar structures $Pn\bar{3}m$ ($NpO_2$) and $Pa\bar{3}$ (from Ref. [133]). In both structures the point site symmetry is $S_6 = C_3 \times i$ of the rhombohedral (trigonal) system. The structures differ by the distribution of the quadupole functions among four crystal sublattices. The $Pn\bar{3}m$ space group has three mirror planes one of which is explicitly shown.

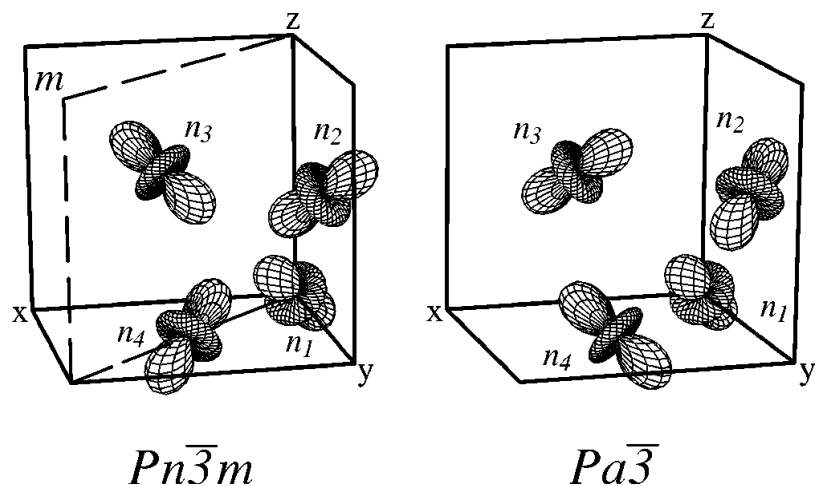

Finally, we would like to remark on the relation between quadrupolar and orbital ordering [137]. If there is only one valence electron at a site, the quadrupolar ordering is equivalent to the orbital one. If however, there are two or more electrons per site then it is impossible to extract only one orbital, because various orbitals of different electrons are mixed up (as for example in the addition of two angular momenta). We believe that the concept of quadrupolar ordering is more correct, because firstly, unlike the orbital, the electron density is an observable quantity, and secondly it is the density which is usually identified with order parameter in the Landau theory of phase transitions [112].

**6.6 Phase transitions with quadrupolar ordering of electron density**

**6.6.1 Hidden structural phase transition in NpO$_2$**

Already in 1953 specific heat capacity measurements in the NpO$_2$ at temperature $T_Q$ =25.5 K revealed a phase transition of unknown nature [19], which become the subject of theoretical and experimental study [138-140]. Since no visible structural distortion was found [139], for many years the phase change in NpO$_2$ was considered isostructural as the γ→α transition in pristine cerium. The disappearance of the localized magnetic moments in the low temperature (ordered) phase further strengthened the analogy. To account for the magnetic effect Santini and Amoretti introduced an octupole magnetic order parameter [140]. The puzzle was solved in 2002, when the phase transition was studied by resonant x-ray scattering experiments performed at the Np $M_{IV}$ and $M_V$ edges [132]. It has turned out that in NpO$_2$ below $T_Q$ there appear superlattice peaks indicative of a long range order of the neptunium quadrupoles. The symmetry of the quadrupolar phase – $Pn\bar{3}m$ is very close to $Pa\bar{3}$ (see Fig. 10). (We recall that earlier the $Fm\bar{3}m \rightarrow Pa\bar{3}$ symmetry lowering was suggested in Ref. [74] for the description of the γ→α phase transition in cerium.) Fig. 10 illustrates that both space symmetries belong to the 3-$q$-AFQ family and differ only by the correspondence between four quadrupolar components and four cubic sublattices [133] (Fig. 10). (The condensation scheme to the $Pn\bar{3}m$-symmetry is given in Ref. [133]). The NpO$_2$ story demonstrates that the structural transition scenario for the γ→α transition in cerium is very probable.

**6.6.2 Quadrupole transitions in CeB$_6$, DyB$_2$C$_2$, UPd$_3$, TmTe and other compounds**

Cerium hexaboride CeB$_6$ is the well known example of cubic compound (the $Pm\bar{3}m$ space symmetry) where a quadrupole transition occurs at temperature $T < T_Q$ = 3.3 K (in zero magnetic field) (see the review of recent works in Ref. [131]). With increasing magnetic field $T_Q$ also increases reaching 10 K at 30 T. Recently the phase transition was studied by means of novel technique of the resonant x-ray spectroscopy [134]. With its help the appearance of the (1/2,1/2,1/2) superlattice diffraction peak in CeB$_6$ [141] and the ($h$/2,$h$/2,$h$/2) reflections ($h$ is odd) in Ce$_{0.7}$La$_{0.3}$B$_6$ [142,143] were detected (in the absence of magnetic field).

There is also a strong indication that in pristine cerium magnetic field can suppress the γ→α phase transition. At least this effect (a complete suppression of the transition in the 56 T magnetic field) was recently found in Ce$_{0.8}$La$_{0.1}$Th$_{0.1}$, where the γ→α transition is shifted to $T$=47 K [144].

Quadrupole transitions were observed in compounds $R$B$_2$C$_2$ ($R$ = Dy, Ho, Tb), in UPd$_3$ and others [131]. The most popular among them is DyB$_2$C$_2$ with the quadrupolar ordering temperature $T_Q$ = 24.7 K (the $P4_2/mnm$ space symmetry). As temperature decreases below $T_Q$ the superlattice peaks ($h$/2,$k$/2,9/2) of reciprocal lattice develop in DyB$_2$C$_2$ [145] and the (1,0,3) reflection in UPd$_3$ [146].

Antiferro-quadrupolar ordering ($T_Q$ = 1.8 K) was found in TmTe [147]. Thulium with one localized 4f-hole at each Tm site is a relatively simple electron system [148]. The peculiarities of the quadrupolar ordering in external magnetic field were studied by neutron spectroscopy which registered the (1/2,1/2,1/2) Bragg reflection of the reciprocal lattice [149].

Note that the appearance of superlattice peaks in all these compounds is in fact an indication of a crystal symmetry lowering. The quadrupolar ordering is nothing else but a low temperature phase transition. A change in structure and symmetry of the local environment of magnetic atom (with an open electron shell) can result in magnetic effects and magnetic ordering. Indeed the quadupole phase transition is very often followed by magnetic ordering at a certain temperature $T_M$ which is lower than $T_Q$. In CeB$_6$ и DyB$_2$C$_2$ that is the antiferromagnetic ordering. The fact that $T_M < T_Q$ implies that it is the quadrupole order parameter which is primary. The scenario when the magnetic order is induced by a change of crystal structure seems crucial and quite general in this situation. Another illustration of this observation is the well known fullerene compound TDAE-C$_{60}$ (this is the only organic compound displaying magnetic order below 16 K). In TDAE-C$_{60}$ two stable phases – α and α' were found differing by the orientation of the C$_{60}$ fullerene molecule. While the α phase is ferromagnetic with the critical temperature $T_M$ = 16 K, its counterpart α' is found nonmagnetic [150].

Experimental aspects of quadrupolar order in TmCu, PrPb$_3$, TbP, DySb, ErAl$_2$, TmGa$_3$, CeMg, CeZn and related phenomena are discussed in reviews [151,152]. According to the authors [152] interactions of quadupole electronic density and induced magnetic moments with crystal lattice (magnetoelastic interactions) can lead to nonlinear magnetization effects or so called metamagnetic transitions.

## 7. Evolution of the electric field gradients at probe $^{111}$Cd nuclei in the cerium lattice

Quadrupolar ordering in a number of compounds (such as NpO$_2$ [132], CeB$_6$, DyB$_2$C$_2$, UPd$_3$ and others [131]) has been detected experimentally by the resonant x-ray diffraction using synchrotron radiation. Nevertheless, the experimental identification of the γ→α transformation in cerium as a structural phase transition with symmetry lowering is a formidable task.

First of all we note weak scattering intensity expected from the quadrupole electron density components of cerium and unusual domain pattern (8 types of domains for $Pa\bar{3}$) imitating the fcc structure. Unfortunately, there are also other technical problems. It is quite difficult to carry out x-ray scattering under pressure. If instead of applying external pressure a cerium sampled is cooled down then the γ→α transition overlaps with other phase transformations (γ→β and β→α). The intermediate β-phase (dhcp) is noncubic and obtained by glide motion of some (111)-planes (see details of this martensitic transition in Sec. 8). Afterwards it is not possible to get rid of the β-phase and at low temperatures the sample is a mixter of β-Ce and α-Ce. Another complication is related to the large volume change at the γ→α transition. If the sample is a single crystal numerous cracks appearing as a result of volume change can simply destroy it. In Refs. [2,44] a by-pass route was used: first α-Ce was reached by compressing the sample to 1 GPa at room temperature, then it was cooled down to liquid-nitrogen temperature and finally the pressure was released. In other experimental works [63,111,135] to get rid of the parasitic β-phase the cerium sample was doped by 7 and even 10 at. % Sc [62].

Notwithstanding all these complications the 3-$q$-AFQ quadrupolar order can be probed practically directly by another experimental method [28,29], which is the time-differential perturbed angular correlation (TDPAC) technique [153,154]. TDPAC belongs to the methods of nuclear spectroscopy where the electric field gradient (EFG) is measured [155] at nuclear probes introduced at sites of crystal lattice. Convenient probes are $^{111}$In/$^{111}$Cd nuclei. Measuring through the hyperfine quadrupolar interactions the electric field gradient at $^{111}$In/$^{111}$Cd nuclei as a function of temperature and pressure [153-155], one obtains information about physical properties of rare earth crystals [156,157]. The hyperfine quadrupole interactions are used in many nuclear methods of solid state physics, for example, in nuclear quadrupole resonance spectroscopy (NQR), in Mössbauer spectroscopy etc. [155]. TDPAC-spectroscopy in particular has the following advantages: 1) it can perform with a small amount of impurity nuclei, 2) in contract to NQR it is not limited to low temperatures. In the best cases TDPAC approaches the accuracy of NQR.

In studies [28,29] the 171-245 keV energy cascade in the $^{111}$Cd nucleus was used, which is populated by the electron capture decay of the $^{111}$In nucleus. In its turn the radioactive $^{111}$In isotope (2.8 day half-life) is produced as a result of the $^{109}$Ag ($\alpha$,2$n$) $^{111}$In reaction by irradiating a silver foil by 32-MeV $\alpha$–beam. Nuclear $^{111}$In probes were introduced in the cerium lattice by melting irradiated silver foil (about 0.1 mg) with cerium powder (500 mg) in a special chamber under the pressure of 8 GPa. The main feature of the TDPAC setup is that it allows us to carry out measurements under external pressure. To the best of our knowledge today it is the only working TDPAC setup with that option. Other experimental details are given in Refs. [158-160].

Measurements were performed on polycrystalline samples of cerium metal (99.98% purity) at room temperature under pressure from the atmospheric value up to 8 GPa. In this range of pressures the cerium sample consecutively transforms to the following phases: $\gamma$, $\alpha$, $\alpha''$, $\alpha'$. The latter phase ($\alpha'$, of the $\alpha$-U type [6]) appears because in the process of melting with a piece of radioactive silver foil the sample was subjected to heat treatment under pressure [8].

Note that before the works [28,29] the TDPAC study of cerium was limited by the $\beta$-phase [161]. The main reason for that is that in the field of the cubic point symmetry (for example at any site of the fcc lattice) there is no quadupole component in the electric crystal field expansion and consequently no electric field gradient. Indeed, it is well known that in the fcc lattice the first nontrivial angular component is given by the cubic spherical harmonic $K_{l=4}(\theta,\varphi)$ with the multipole index $l=4$ (see for example Ref. [114]). This implies that all components with $l=1,2,3$ are zero. Since the tensor $V_{ij}$ of EFG is described by the multipole index $l=2$, it follows that $V_{ij}=0$, and the TDPAC method is ineffective, because it gives no information on the cubic crystal field.

The disappearance of the electric field gradient in the fcc lattice can be observed experimentally. Such data (at the $^{111}$Cd probe nuclei in the thallium lattice) are reported in the work of da Jornada and Zawislak [162]. At atmospheric pressure the thallium lattice is the hexagonal close packed structure (the $P6_3/mmc$ space group), in which the cadmium probes register the electric field gradient of 8.0(4) MHz [162]. With increasing pressure, the quadrupole frequency $\nu_Q$ (or EFG) slowly and monotonically decreases down to 6.1(5) MHz (at $P=3.5$ GPa), but when at slightly elevated pressure ($P=3.7$ GPa) a structural phase transition to the fcc structure occurs and the electric field gradient drops to zero (0.0(4) MHz) [162].

This effect offers a unique way to check the isostructural feature of the $\gamma\rightarrow\alpha$ phase transition in cerium. If the $\gamma\rightarrow\alpha$ transformation is truly isostructual, that is if the $\gamma$- and $\alpha$- phases are characterized by the same cubic point group, then in both phases the gradient should be zero: $V_{ij}=0$ (in practice $V_{ij}\approx 0$). However, it has turned out that $V_{ij}\approx 0$ only in the $\gamma$-phase, whereas in the $\alpha$-phase $V_{ij}\neq 0$ [28,29]. This implies the appearance of quadrupole component of electron density at the $^{111}$Cd probe sites and an effective crystal symmetry lowering of the cerium sample. Below we discuss the results of experiments [28,29] in more detail.

**Figure 11.**
Pressure dependence (at room temperature) of the nuclear quadrupole frequency ($\nu_Q$, left axis) and the electric field gradient ($V_{zz}$, right axis) at $^{111}$Cd probe nuclei in cerium lattice crystal sites. The results of TDPAC measurements are from Refs. [28,29], data for β-Ce are from Ref. [161].

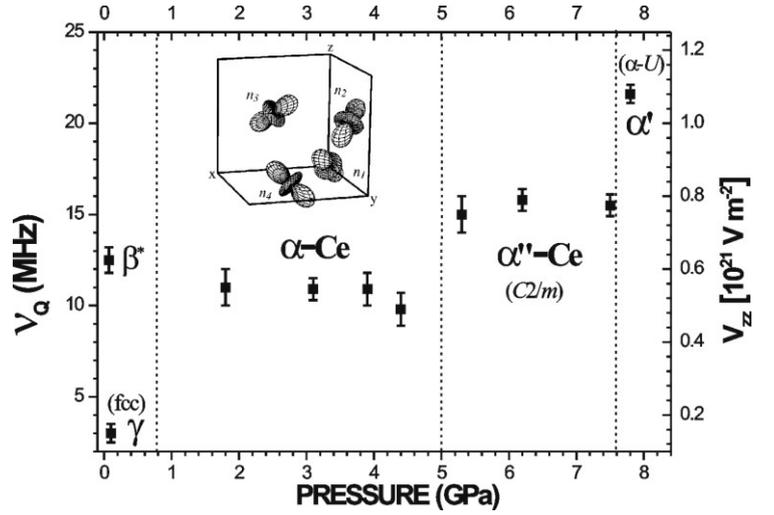

In the γ-phase at atmospheric pressure $\nu_Q(\gamma)$ = 3 MHz [28,161]. This is practically a background level (Fig. 11 and Table 7). The value is relatively large and later we will discuss it. From the general considerations one can expect approximately the same frequency (EFG) in α-Ce, whereas it has been found that $\nu_Q(\alpha)$ = 11 MHz which is 3.7 times higher. The electric field gradient measured under pressure from the atmospheric value to 8 GPa [28] is reproduced in Fig. 11 and Table 7. Notice that EFG in α-Ce is comparable with EFG in other noncubic phases (β- and α"-phases), which border α-Ce in the *P-T* phase diagram. Thus, the quadrupole electron density component in α-Ce is approximately the same as in *other noncubic phases*. This experimental finding unambiguously indicates that the γ→α transition in cerium is not isostructural, and the symmetry of the α-phase differs from the γ-phase ($Fm\bar{3}m$). Earlier, a 3-*q*-AFQ order was predicted in the quadrupole model [74,75]. Unfortunately, TDPAC spectroscopy cannot determine the exact space group for the α-phase. Possible candidates here are $Pa\bar{3}$ and $Pn\bar{3}m$.

Fig. 11 indicates that in addition to the γ→α transformation there are two other jumps of EFG at pressures 5 and 7.5 GPa. They correspond to two other phase changes: α→α", α"→α' (see Sec. 9). In the case of α'-Ce a nonzero asymmetry parameter of EFG, η = 0.52, was found (Table 7), which is characteristic of the α-U structures. (In particular, in the α-phase of uranium η = 1 [163].)

**Table 7.** The electric field gradient $V_{zz}$ (the quadrupolar frequency $\nu_Q$) and the asymmetry parameter η at the probe $^{111}$Cd nuclei in cerium lattice at room temperature under pressure (*P*). (Data from Refs. [28,29]).

| *P*, GPa | $\nu_Q$, MHz | η | $V_{zz}$, $10^{21}$ V m$^{-2}$ |
|---|---|---|---|
| 0 | 3.0(5) | 0 | 0.15(3) |
| 1.8 | 11(1) | 0 | 0.55(5) |
| 3.1 | 10.9(6) | 0 | 0.54(3) |
| 3.9 | 10.9(9) | 0 | 0.54(4) |
| 4.4 | 9.8(9) | 0 | 0.49(4) |
| 5.3 | 15(1) | 0 | 0.75(5) |
| 6.2 | 15.8(6) | 0 | 0.79(3) |
| 7.8 | 21.6(5) | 0.52(5) | 1.08(3) |

In publication of the work [28] a reviewer pointed out that almost all TDPAC spectra $R(t)$ were obtained during the observation time $t < T$, where $T$ is the period of TDPAC-oscillations, which reduces the accuracy of measurements. This was certainly true, and even more so, some additional technical difficulties were caused by applying external pressure [160,164]. However, the same situation ($t < T$) was observed at the $^{111}$Cd nuclei in thallium, but the hcp→fcc transition there was clearly detected [162]. The $t < T$ relation is caused by a low quadrupole frequency: 6-8 MHz in Tl [162] and 10 MHz in Ce [28,161]. As pointed out in Ref. [28] in the case of cerium measurements were carried out with several samples. The EFG jumps were observed both with increasing pressure (γ→α) and decreasing it (α→γ) so that from our TDPAC experience it certainly follows that the effect of quadupole ordering in α-Ce does take place.

There also was a concern caused by a relatively large value of EFG in the γ-phase (3 MHz). This is not a simple question. It should be noted that a small nonzero gradient at $^{111}$Cd probes is detected for all cubic crystals, but in cerium it is somewhat larger than a typical value. However, the 3 MHz gradient and the TDPAC spectrum of γ-Ce of Ref. [28] practically coincide with the spectrum and $\nu_Q$ found by Forker et al. [161]. On the other hand, in a number of cerium samples of Ref. [28] a smaller value of $\nu_Q(\gamma)$ was detected, so that $\nu_Q(\gamma)$ depends on the sample quality and prehistory. Below we mention the factors which could produce a nonzero value: 1) the structure of domains and polycrystalite grains, 2) other phase nucleations (for example, β-Ce), 3) small local strains caused by the replacement of Ce with Cd. Such replacements induce small electric quadupole moments which in principle can cause their local freezing [165]. The latter factor is inherent in the experimental method and cannot be excluded. Summarizing, $\nu_Q(\gamma)$ should be considered as a background level which we have to subtract from EFG of other phases. In any case, almost fourfold increase of $\nu_Q$ at the γ→α phase transition cannot be explained within the isostructural paradigm.

## 8. Peculiarities of electronic structure of β-Ce and δ-Ce

### 8.1 β-Ce (double hexagonal close packed structure, dhcp)

The transformation of the fcc structure of γ-cerium to the dhcp structure of β-cerium is martensitic. It depends on the purity and the grain size, proceeds very slowly and never fully completes. The amount of the β-phase in a sample depends on 1) the cooling rate, 2) the temperature to which the sample is cooled, and 3) the number of cooling-heating cycles. The start of the transformation on cooling (i.e. γ→β) is in the temperature range 240-290 K, on heating (β→γ) at 373-451 K [2]. A 20-year study of Gschneidner et al. puts the low boundary of the transformation at 283 K [166]. Interestingly, the γ→β transition is accompanied by a small volume increase (1.2%). As a result, the specific volume of β-Ce is slightly larger than the volume of the δ-phase (0.2%) and is the largest among all cerium allotropies (see Table 2). In β-Ce the closest distances between the cerium atoms belonging to neighboring hexagonal planes slightly decreases (0.05%), while distances within the planes increases (0.88%) in comparison with the closest distances in γ-Ce (see Table 2). (The lattice constants are taken from Ref. [16].) More details about the transformation can be found in the review of Koskenmaki and Gschneidner [2].

McHargue and Yakel proposed a dislocation model for the mechanism of the γ→β transformation in cerium [167]. It is well known that in the fcc lattice the closed packed (111) planes (perpendicular to the [111] vector) form the layering sequence of the type …ABCABC… (Fig. 12). According to the mechanism of McHargue and Yakel, two of these planes, say C and A, as a result of two glide motions transform to the planes A and C, respectively. Next two planes, which are A and B, remain in their positions (see Fig. 12). Two following planes C and A again undergo glide displacements and become A and C and etc. The resultant packing sequence changes to …ABACABAC…, which corresponds to the dhcp layering (see Fig. 12). Note that in the dhcp structure there are two nonequivalent cerium sites, so the transformation certainly involves a symmetry lowering. It is often said that the sites of the planes of the A type are in the "cubic" (i.e. CAB) environment, because the neighboring planes are of C- and B-type, while the cerium sites in C- and B- planes (ACA or ABA) are in the hexagonal environment. One should keep in mind though that since in the β-phase the ratio $c/2a = 1.611$ differs from the cubic one ($c/a = \sqrt{8/3} \approx 1.633$), the exact point symmetry there is not cubic. In particular, at the sites A there is only one threefold axis of symmetry. Notice also that the planes C- and B- can be considered as the mirror symmetry planes, and the dhcp structure can be viewed as an ideal twinned crystal. The martensitic character of the transformation is supported by the observation that in the range 196-500 K plastic deformations assist the γ→β transformation. Obviously, the appearance of numerous glide motions lead to appreciable microdeformations in the crystal grains.

Although the rate of the γ→β change is very slow the transformation cannot be avoided even on fast cooling (quenching). This hinders the study of another phase transition, γ-Ce (fcc) → α-Ce (cubic), which is of fundamental interest. From the viewpoint of the γ→α transition the phase transformation to the β-phase at low temperatures is a parasitic process which should be completely suppressed. As a solution to this problem in a number of works [62,63,111,135] cerium samples were doped by scandium: $Ce_{1-x}Sc_x$, where $x=0.07$ [63,111,135] or $x=0.1$ [62]. Atoms of the trivalent scandium have a smaller metallic radius (1.62 Å), which results in a smaller lattice constant of the $Ce_{1-x}Sc_x$ fcc crystal. This is equivalent to an effective external pressure $P_{eff}$ applied to shift the γ→α transition to a phase diagram region where the intermediate β-phase is absent. However, the scandium amount is not negligible and these samples cannot be considered as pristine cerium. Such cerium-scandium compounds, $Ce_{1-x}Sc_x$, are not always suitable for studies of subtle effects.

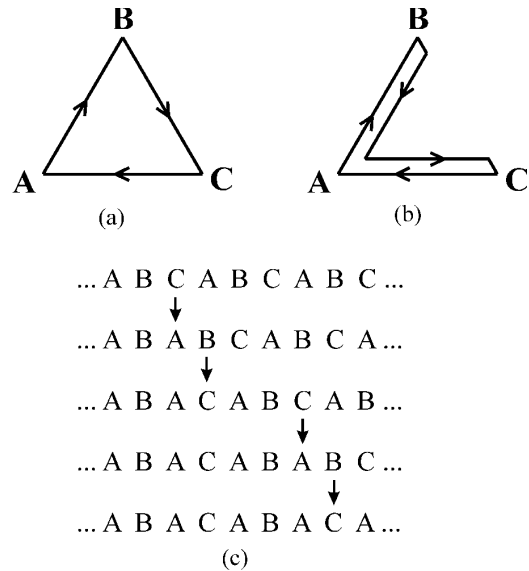

**Figure 12.**
The formation of the double hexagonal structure (dhcp, β-Ce) from the face centered cubic lattice (fcc, γ-Ce). (a) the sequence of the (111)-layers in the fcc lattice (ABCABC); (b) the sequence of the (111)-layers in the dhcp lattice (ABACABAC); (c) glide planes transforming the fcc sequence (top) to the dhcp sequence (bottom).

Recently, refined data on β-Ce and γ-Ce were reported [166,168-170]. The electron specific heat (γ) is $(7.0 \pm 0.1)10^{-3}$ J mole$^{-1}$ K$^{-2}$ for β-Ce and $(6.2 \pm 0.8)10^{-3}$ J mole$^{-1}$ K$^{-2}$ for γ-Ce. Thus earlier values [2] were decreased. In Ref. [166] from the latent heat of the β→γ transformation at 420 K an entropy change of $\Delta S^{\beta-\gamma} = 0.05 k_B$ per cerium atom was obtained. There are two contributions to the entropy change, one is from the lattice, $\Delta S^{\beta-\gamma}_{vib} = (0.09 \pm 0.05) k_B$, and the other is electronic, $\Delta S^{\beta-\gamma}_{el} = -(0.04 \pm 0.05) k_B$ [169]. The result shows that the lattice term causing the softening of the phonon spectrum in γ-Ce is principal, while the electronic contribution acts in the opposite direction [169] destabilizing the β→γ transition.

Another anomaly of the β-phase which we have mentioned earlier is that its volume is larger than the volume of any other cerium phase (Table 2). This is an indication that the very existence of the β-phase is related with its intermediate role in respect to the γ–α transition. β-Ce is stabilized by the polycrystalline grains, interphase boundaries, twinings and other defects. Indeed, in a single crystal the volume change at the γ→α transition accompanied by considerable strains cleaves the crystal. In polycrystals the same elastic strains between grains and domains of the α-phase move some of the (111)-planes to the intermediate β-structure. It follows then that any defects, for example in area between domains and other phases, are partially compensated and softened by the growth of grains of the β-phase with large volume per atom. This observation explains the well known fact that the amount of β-Ce on cooling from γ-Ce to α-Ce is 20-30% [2], and can be increased by multiple γ-Ce → α-Ce → γ-Ce cycling.

Because of the nature of the β-phase, the growth of a β-Ce single crystal seems impossible. Nevertheless, it has turned out that a single crystal β-phase can be obtained by the molecular epitaxy method on Nd surface, which stabilizes it down to low temperatures ($T = 2$ K) [171]. In Ref. [171] such epitaxial β-phase containing 60 layers of $Ce_{30}/Nd_{10}$ was studied in detail. The lattice parameters are $a$=3.664(3) Å, $c$=2.970(3) Å (in niobium $c$=2.949(3) Å) implying a small compression within the hexagonal planes (0.5%) and a small expansion between the planes (0.2%) in comparison with the bulk β-phase [16]. Unlike the bulk phase the epitaxial β-phase (e-β-Ce) does not transform to α-Ce. Moreover, the epitaxial γ-Ce on cooling does not undergo the transition to e-β-Ce. Instead, at $T$=120 K e-γ-Ce ($Ce_{30}/Nd_{10}$) transforms to a phase with the samarium structure which is very typical of the 5d metals [172]. This is another indication of very important relation between β-Ce and the γ–α transition in polycrystalline sample underlining the decisive role of the elastic strains for the stabilization of the β-phase.

The neutron diffraction measurements of the magnetic order in e-β-Ce (at $T$=7 K [173]) confirmed the results reported earlier for the $Ce_xY_{1-x}$ compounds [174,175]. The antiferromagnetism is related with the transverse order (along one of the $a$ directions of the hexagonal lattice) of the $0.1\mu_B$ magnetic moments. (In $Ce_xY_{1-x}$ the magnetic moments are larger [174,175].) The vector of the sign alternation (the [1/2,0,0] point of the Brillouin zone) is perpendicular to the magnetic moments (Fig. 13). The antiferromagnetic ground state of β-Ce (with the $1.1\mu_B$ magnetic moments) is reproduced by the band structure calculations only in the LDA+U approximation, while the standard density functional calculations (LDA or GGA approximation) predict the ferromagnetic ground state [175].

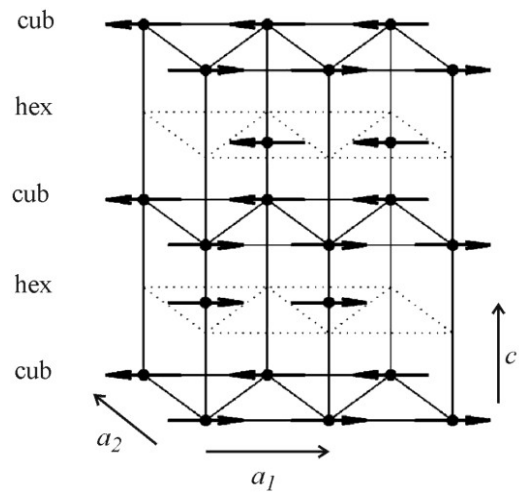

**Figure 13.**
Antiferromagnetic transverse ordering in β-Ce. (Data for the epitaxial phase of β-Ce, Ref. [171] and $Ce_{0.75}Y_{0.25}$, Ref. [173].)

## 8.2 δ-Ce (body centered cubic structure, bcc)

In contrast to closed packed fcc and dhcp phases, the body centered cubic lattice (bcc) is an open structure with large volume in the interstitial region. Bcc structures are characteristic of high temperature phases of transitional elements with partially filled d-shell. As a rule, on cooling the phases give place to closed packed structures but this instability rapidly disappears with increasing d-shell occupation. Bcc phases are also found in rare earths but there they exist in a narrow temperature range of about 10 K. In cerium this temperature region is much larger (69 K) which also characterizes it as a d-element. (The important contribution of d-electrons to the metal bond of cerium is discussed in detail in Sec. 2.) Thus, on cooling from the cerium melt the sample undergoes the following transformations: Ce (melt)–(1068 K)→ δ-Ce –(999 K)→ γ-Ce. Although the δ–γ transition is well defined (unlike the γ–β change) the cooling procedure nevertheless leads to sample contamination by other phases (β,α). To avoid it in [82] a single crystal of δ-Ce was grown *in situ* (on cooling below 999 K) in the argon atmosphere, and after this its phonon spectrum was measured by inelastic neutron scattering at $T$=1036 K. The disperse curves of δ-Ce then were fit according to the standard Born-von Kármán force model with five nearest shells (Table 8). As a result the elastic constants $c_{11}$=2.20(7), $c_{12}$=1.71(8) and $c_{44}$=1.47(3), in $10^{10}$ N/m$^2$ (or $10^{11}$ dyne/cm$^2$) were obtained from the direct fitting of the phonon spectrum, or alternatively $c_{11}$=2.23, $c_{12}$=1.73 and $c_{44}$=1.62, in $10^{10}$ N/m$^2$ ($10^{11}$ dyne/cm$^2$) from the force constants.

The important feature of the phonon spectrum of δ-Ce is a pronounced minimum of longitudinal vibrations at ξ=2/3 (or L2/3) along the [ξ,ξ,ξ] direction, and also low lying transverse vibrations along $T_1$[ξ,ξ,0] and $T_2$[ξ,ξ,2ξ]. These anomalies are an indication of instability towards the closed packed structures (fcc, hcp, dhcp). In addition, the phonon spectrum is considerably damped especially at the Brillouin zone boundaries where the halfwidth is of the same order as the phonon frequency. The large ratio of two shear constants, $A=c_{44}/c'$=5.6 (see Sec. 4.1 and Eq. (1)), which is only 2.8 for γ-Ce, implies the large anisotropy of the crystal potential and its instability to tetragonal distortions. The estimated Debye temperature is 95.2 K, the mean square displacement of a cerium atom is 0.119 Å$^2$. In general the phonon spectrum of δ-Ce is noticeably softer than in γ-Ce, especially in the frequency range ν < 1 THz, which causes the 0.45$k_B$ excess of the lattice entropy per atom. (The total entropy change at the γ→δ transition including the electronic contribution is 0.35$k_B$.)

**Table 8.**

Force constants $\Phi_{\alpha\beta}(\vec{R}_n)$ (in $10^2$ n/m$^2$ or $10^3$ dynes/cm$^2$) of δ-Ce in the Born-von Kármán model with 5 nearest coordination shells obtained by fitting the phonon spectrum at 1036 K [82]. $N$ is the number of atoms in shell, $\alpha$ ($\beta$) = $x, y, z$.

|   | $\vec{R}_n$ | $N$ | $\Phi_{\alpha\beta}(\vec{R}_n)$ |
|---|---|---|---|
| 1 | $a$(1/2,1/2,1/2) | 8 | $\Phi_{xx}$ = 2.945, $\Phi_{xy}$ = 3.563 |
| 2 | $a$(1,0,0) | 6 | $\Phi_{xx}$ = 2.313, $\Phi_{yy}$ = –0.856 |
| 3 | $a$(1,1,0) | 12 | $\Phi_{xx}$ = 0.048, $\Phi_{zz}$ = 0.171, $\Phi_{xy}$ = 0.346 |
| 4 | $a$(3/2,1/2,1/2) | 24 | $\Phi_{xx}$ = –0.021, $\Phi_{yy}$ = 0.187, $\Phi_{xy}$ = –0.018, $\Phi_{yz}$ = –0.193 |
| 5 | $a$(1,1,1) | 8 | $\Phi_{xx}$ = –0.262, $\Phi_{xy}$ = –0.120 |

Therefore, in contrast to the fcc structure of γ-Ce, the bcc structure of δ-Ce is stabilized by the entropy lattice contribution.

Note that lattice properties and the phonon spectrum of δ-Ce are very similar to properties of the homologue bcc phases of β-Sc and γ-La. As δ-Ce both β-Sc and γ-La transform to closed packed structures (β-Sc to hcp, γ-La to fcc). As in cerium the metallic bond in Sc and La is formed by one d- and two s-electrons (see Sec. 2). This is a further indication that 4f electrons of cerium do not participate directly in chemical bonding [82] (Sec. 2).

## 9. High pressure phases of cerium (α'', α' and ε)

### 9.1 Identification of α''-Ce and α'-Ce

At room temperature and pressure $P \approx 5$-$5.5$ GPa α-Ce transforms to an allotropic modification which was called α'-Ce. The identification of the α' crystal structure became "one of the most controversial subjects concerning high-pressure phases" [176]. First, α'-Ce was considered as the fcc structure with a lattice constant which is 4% smaller than in α-Ce [177]. Then, hcp structure [178] and finally C-centered orthorhombic [10] (the α-U structure) (Fig. 14) were reported for α'-Ce. In a series of x-ray diffraction studies Zachariasen and Ellinger have excluded fcc and hcp as erroneous but simultaneously found a new cerium phase which they called α'' [10]. This monoclinic α''-phase has the $I2/m$ space symmetry (with the lattice constants $a = 4.762$, $b = 3.170$, $c = 3.169$ Å, $\beta = 91.73°$) which can be viewed as a distorted fcc structure. It was first described as a metastable phase: upon releasing pressure it transforms to α-Ce, on increasing pressure to α'-Ce. Then, another monoclinic phase of the $C2/m$ space symmetry was reported [7]. Thus, two α''-phases appear: α''-Ce(I) and α''-Ce(II). Later however it was demonstrated that both phases are identical and have the same $C2/m$ space symmetry [8]. It has four atoms in the primitive unit cell and can be considered as a superstructure formed by doubling the $I2/m$ structure with two atoms in its unit cell (Fig. 15). The problems with the phase identification arose because x-ray patterns were strongly affected by sample orientation [8].

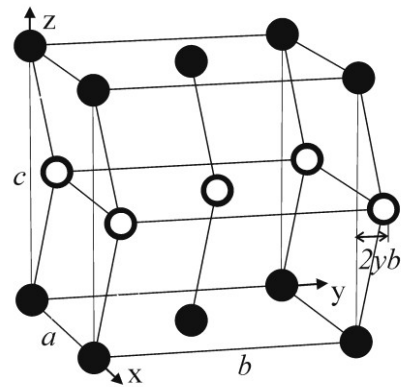

**Figure 14.**

The base centered (C) orthorhombic primitive unit cell (α'-Ce, the *Cmcm* space group, the α-uranium structure). It becomes the fcc structure if $a=b=c$, $y=0.25$.

**Figure 15.**
The relation between the monoclinic primitive unit cell (α''-Ce, the $C2/m$ space symmetry) and the fcc lattice (γ-Ce). (Refs. [8,148].)

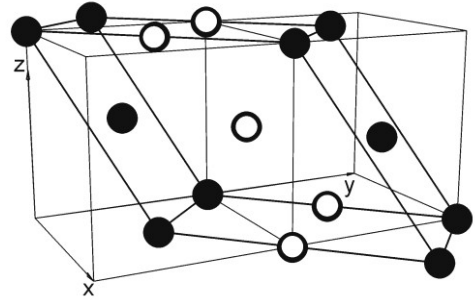

Thus, it has been established that the sequence of the cerium phase transformations under pressure is the following:

$$\gamma\text{-Ce (fcc)} \rightarrow \alpha\text{-Ce (cubic)} \rightarrow \alpha'\text{-Ce }(\alpha\text{-U}) \text{ or } \alpha''\text{-Ce }(C2/m) \rightarrow \varepsilon\text{-Ce (bct)} \qquad (10)$$

However, in this sequence it is not clear which phase – α'-Ce (α-U) or α''-Ce ($C2/m$) – is stable and which is metastable under pressure 5 to 12 GPa. This question has been debated for long time and apparently was closed by the study of McMahon and Nelmes [8]. α'-Ce is slightly denser (0.11(8)%) and the nearest distance between cerium atoms there is 0.1 Å smaller than in α''-Ce. This indicates that the α-U structure should be more energetically favorable than $C2/m$. However, the α-U structure is hexagonal-like, the transition to which invokes relatively large displacements accompanied by additional energy losses whereas the transformation to the monoclinic phase causes only small atomic displacements from the cubic positions. In Ref. [8] the following experiment is described. Starting at room temperature from almost pure α'-Ce (α-U) at P = 7 GPa, pressure was reduced to 2.5 GPa. When transformation to the cubic α-phase was complete pressure was increased to the initial value (7 GPa). It was then found that the sample was almost entirely in the α''-phase ($C2/m$).

Glide motions of planes are responsible for other peculiarities of the transformation to α'-Ce (α-U). It has turned out that the α-Ce (cubic) → α'-Ce (α-U) transition occurs from a very few centers of the α-phase so the phase change apparently depends on grain size. It is also possible that the α→α' transformation requires a minimum critical size of grains. The authors of Ref. [8] pointed out that this observation can explain different branches of the phase sequence, that is which path: α-Ce → α'-Ce (α-U) or α-Ce → α''-Ce ($C2/m$) occurs in a cerium sample depends on the method of sample preparation. In "cold-worked" samples which were not annealed at high temperature and pressure the α-Ce → α''-Ce ($C2/m$) transition takes place. If however, sample is preliminary heated it supposedly contains nuclei of the α'-phase (α-U) and transforms to that phase. Interestingly, to some extent the trend is opposite to that in the γ → β transition which is enhanced with lowering temperature. Thus, McMahon and Nelmes [8] gave convincing arguments that it is α''-Ce which is thermodynamically stable. The opposite viewpoint is represented by the authors of Refs. [35,25]. They argue that if at pressure 5 GPa these two phases (α' and α'') coexist, then at pressures above 7 GPa the α''-Ce fraction disappears while α'-Ce is present up to 13 GPa. In Ref. [25] one can find other arguments in favor of metastability of α''-Ce.

The last transformation in (10) occurs when P > 12 GPa (start at P = 12.5 GPa, end at P = 17.7 GPa.) The structure and symmetry of ε-Ce are well established. It is the body centered tetragonal lattice [9] with the lattice parameters given in Table 1. However, in this range of pressures and temperatures various phases

(α, α', α" and ε) can be present in a cerium sample and therefore several phase transformations with large hysteresis effect are often detected [25]. Complex behavior of such a phase mixture was studied by Tsiok and Khvostantsev in Ref. [25] who concluded that phase composition depends not only on temperature and pressure but even on the trajectory in the *P-T* plane, which leads to a chosen point of the *P-T* diagram (the region of phase ambiguity in Fig. 1). For the first time arched phase (α, α', α" and ε) boundaries were mentioned in the work of Antonova et al., Ref. [23].

Note also that Tsiok and Khvostantsev, based on resistance measurements, reported the direct transformation from ε-Ce to α-Ce [25]. The transition in the direction α→ε was established by the x-ray diffraction [26]. The slope of the transition boundary between α-Ce and ε-Ce is positive (Fig. 1), but its values in [25] and [26] are quite different.

### 9.2 Condensation schemes of symmetry lowering in high pressure phases

From the group-theoretical approach the symmetry lowering from fcc (γ-Ce) to the monoclinic α"-Ce (*C2/m*) is driven by the condensation of order parameter at the *L* point of the Brillouin zone [179]. The corresponding wave vector $\vec{q}_L$ ($k^9$ in Kovalev's notation [116]) has four rays $\vec{q}_L^i$ (*i*=1-4), where each ray is invariant to all symmetry operations of the small group $\bar{3}m$ ($D_{3d}$), with the three-fold axis as the main symmetry element. On the basis of two dimensional ($E_g$) representation of the small group (with the basis density functions $\rho_1^i$ and $\rho_2^i$) one obtains 8 basis functions of the irreducible representation $L_3^+$ ($k^9$) [115,116]. As shown in Ref. [148] the condensation of the first function ($\rho_1^i$) of a ray leads to the *C2/c* space symmetry, while the condensation of the second ($\rho_1^i$) to *C2/m*. Therefore, the condensation scheme to α"-Ce reads as

$$Fm\bar{3}m : L_3^+[\rho_2(\vec{q}_L) = \rho] \rightarrow C2/m.$$

There are 12 domains of *C2/m*.

Transformation to α'-Ce (α-U) with the *Cmcm* symmetry ivolves the irreducible representation $X_5^+$ [115] ($k^{10}$ in Kovalev's notation [116]) at the *X* point of the Brillouin zone. In that case there are three rays $\vec{q}_X^j$ (*j*=1-3), and the small group is $\bar{4}m$ ($D_{4d}$). Taking as basis functions of two dimensional representation ($E_g$) the density components $\rho_1^j$ and $\rho_2^j$ one gets the six dimensional irreducible representation $X_5^+$. Condensation of three out of its six components gives a cubic symmetry ($Pn\bar{3}m$ or $Pa\bar{3}$), which most likely realized in α-Ce [74] (see Sec. 6). If however the condensation involves only one component ($\rho_1^j$ or $\rho_2^j$) of one ray, then the symmetry lowers to *Cmcm* (the α-U structure [5,6]), which is realized in α'-Ce:

$$Fm\bar{3}m : X_5^+[\rho_1(\vec{q}_X) = \rho] \rightarrow Cmcm.$$

This symmetry change results in 6 different domains. Condensation schemes for cerium phases are summarized in Table 9.

Interesting group-theoretical relation between the space symmetries α"-Ce (*C2/m*) and α'-Ce (*Cmcm*) is presented in Ref. [180]. The authors proposed to consider the body centered cubic lattice (bcc) as a parent structure, from which through displacive mechanisms and symmetry lowering all other phases can be deduced. The displacive transformations are divided in two groups: (1) variants of the Burgers mechanism transforming bcc to hcp, dhcp and 9R structures [181], and (2) variants of the Bain deformations transforming bcc to fcc or bct [182]. The transformation from bcc to the orthorhombic α'-phase (α-U, or *Cmcm*) is described by a condensation of one out of six components at the $N_b$ $(2\pi/a)(1/2,1/2,0)$ point of the Brillouin zone of the bcc lattice. The condensation can occur via any component, each of the variants corresponds to one of six domains. If now a Bain deformation is applied transforming bcc to fcc then the six former equivalent domains become nonequivalent. Some of the domains still correspond to α'-Ce (with the *N* point becoming the *X* point of the Brillouin zone of the fcc lattice) while the other domains correspond to the monoclinic α"-phase (with the *N* point becoming the *L* point of the Brillouin zone of the fcc lattice). This mechanism gives a nontrivial group-theoretical relation between α'-Ce and α"-Ce and explains a low energetic barrier between the two phases.

**Table 9.**

Symmetry lowering (order parameter condensation schemes) and space group relations between the (high symmetry) fcc structure of γ-Ce and phases of low symmetry. "Rays" and "components of one ray" here stand for rays and the components of one ray which are involved in the corresponding condensation scheme.

| Phase | irreducible representation (BZ point) involved [115] | number of rays | number of the components of one ray | qudrupole components' symmetry in one ray | space symmetry |
|---|---|---|---|---|---|
| α | $X_5^+$ | 3 | 1 | $T_{2g}$ | $Pn\bar{3}m$ or $Pa\bar{3}$ |
| α' | $X_5^+$ | 1 | 1 | $T_{2g}$ | *Cmcm* |
| α" | $L_3^+$ | 1 | 1 | $T_{2g}$, $E_g$ | *C2/m* |
| ε | $X_2^+$ | 1 | 1 | $E_g$ | *I4/mmm* |

**10. Conclusions**

We have reviewed phase transitions in metallic cerium taking into account recent experimental data including high pressure measurements [8,25,26]. Much attention has been given to the γ→α phase transformation and in particular to its elastic anomalies [69,80,27], and novel experimental results [28,29] (TDPAC-spectroscopy), which have revealed a quadupole electron density component in α-Ce. This is an unambiguous indication that the γ→α transition is a hidden structural transition with space symmetry lowering [73-75].

Of much interest are also experimental data on the epitaxial β-phase of Ce grown on niobium surface [171]. On cooling instead of transforming to α-Ce this epitaxial β-phase undergoes a transition to samarium-type crystal structure characteristic of 5d-metals. These and other observations suggest that the appearance of the bulk β-phase in the *P-T* region between γ- and α-Ce is caused by specific effects.

We have also critically discussed recent computational methods and calculations of the γ→α phase transition, considered in detail the quadrupole model [74,37,186], the peculiarities of the electron structure of atomic cerium and the simplest chemical bond in the cerium dimer.

We are most grateful to K.H. Michel, S.M. Stishov, V.V. Brazhkin and other colleges for valuable suggestions and fruitful discussion.
The work is financially supported by RFBR, grant № 11-02-00029.